\documentclass[twocolumn, 10pt]{IEEEtran}

\usepackage{etoolbox}
\makeatletter
\patchcmd{\@begintheorem}{\textit}{\textbf}
\makeatother

\usepackage{hyperref} 
\usepackage{graphicx}
\usepackage{amssymb}
\usepackage{amsmath}
\usepackage{mathtools}
\usepackage{cite}
\usepackage{stfloats}
\usepackage{subfigure}
\usepackage{psfrag}
\usepackage[mathscr]{euscript}
\usepackage{acronym}  
\usepackage{booktabs} 
\usepackage[table]{xcolor}

\acrodef{NTN}{non-terrestrial network}
\acrodef{3GPP}{3rd generation partnership project}
\acrodef{HAPS}{high altitude platform systems}
\acrodef{LAP}{low altitude platform}

\acrodef{BS}{base station}
\acrodef{UE}{user equipment}
\acrodef{UAV}{unmanned aerial vehicle}
\acrodef{LoS}{line-of-sight}
\acrodef{NLoS}{non-line-of-sight}
\acrodef{IoT}{Internet-of-Things}
\acrodef{GU}{ground user}
\acrodef{AU}{aerial user}
\acrodef{SNR}{signal-to-noise ratio}
\acrodef{SIR}{signal-to-noise ratio}
\acrodef{SINR}{signal-to-interference-plus-noise ratio}
\acrodef{HPPP}{homogeneous Poisson point process}
\acrodef{ES-BS}{exclusive-service \ac{BS}}
\acrodef{IS-BS}{inclusive-service \ac{BS}}
\acrodef{PDF}{probability distribution function}
\acrodef{CDF}{cumulative distribution function}
\acrodef{CCDF}{complementary cumulative distribution function}
\acrodef{NOMA}{non-orthogonal multiple access}
\acrodef{LTE}{long term evolution}
\acrodef{MIMO}{multiple-input multiple-output}
\acrodef{AF}{amplify-and-forward}
\acrodef{DF}{decode-and-forward}
\acrodef{PGFL}{probability generating functional}
\acrodef{CoMP}{coordinated multi-point}
\acrodef{MHCPP}{Matérn Hardcore Point Processes}
\acrodef{GBS}{\ac{BS} for \acp{GU}}
\acrodef{ABS}{\ac{BS} for \acp{AU}}

\usepackage{color}
\usepackage{dsfont}
\usepackage{bbm}

\newtheorem{theorem}{Theorem}
\newtheorem{lemma}{Lemma}
\newtheorem{corollary}{Corollary}

\newtheorem{remark}{Remark}

\newcommand{\noise}{\sigma^2}
\newcommand{\target}{\gamma_\text{t}}
\newcommand{\minpower}{\eta}
\newcommand{\tpower}{P_\text{t}}
\newcommand{\userratio}{\rho_{\text{G}}}
\newcommand{\bsratio}{\rho_{\text{B},\text{G}}}

\newcommand{\hpppi}{\Phi_{\text{U},i}}
\newcommand{\hpppBS}[2]{\Phi^{#2}_{\text{B},#1}}
\newcommand{\hpppI}[1]{	\Phi_{\text{I},#1}		}
\newcommand{\hBS}{h_\text{B}}
\newcommand{\hAU}{h_\text{A}}
\newcommand{\hGU}{h_\text{G}}
\newcommand{\hi}{h_{k}}
\newcommand{\angledb}{\theta_\text{3dB}}
\newcommand{\angleth}{\theta_{\text{th}}}
\newcommand{\eangle}{\theta\left( \hd \right) }
\newcommand{\tilt}{	\theta_{\text{t}}	}
\newcommand{\tilti}{	\theta^s_{\text{t,}i}		}
\newcommand{\tiltg}{	\theta^s_{\text{t,G}}		}
\newcommand{\tilta}{	\theta^s_{\text{t,A}}		}
\newcommand{\tiltl}{\theta^s_{\text{t,}l}}
\newcommand{\tiltIS}{\theta^\text{IS}_{\text{t,O}} }

\newcommand{\tiltgIS}{	\theta^\text{IS}_{\text{t,G}}		}
\newcommand{\tiltaIS}{	\theta^\text{IS}_{\text{t,A}}		}
\newcommand{\tiltgES}{	\theta^\text{ES}_{\text{t,G}}		}
\newcommand{\tiltaES}{	\theta^\text{ES}_{\text{t,A}}		}

\newcommand{\optiltg}{	\theta^*_{\text{t,G}}		}
\newcommand{\optilta}{	\theta^*_{\text{t,A}}		}
\newcommand{\optiltIS}{	\theta^*_{\text{t,O}} }
\newcommand{\optilti}{	\theta^*_{\text{t,}i}		}
\newcommand{\interference}[2]{I^{#1}_{#2}}
\newcommand{\hd}{r_{k,\mathbf{x}}}
\newcommand{\shd}{r_{k,\servingBS}}
\newcommand{\ld}[1]{	l_{v\hspace{-0.5mm}}\left( #1 \right)  }
\newcommand{\upDistance}[1]{r_{#1}^{\text{ub}}\left( \tilt \right) }
\newcommand{\loDistance}[1]{r_{#1}^{\text{lb}}\left( \tilt \right) }

\newcommand{\bound}[1]{b_{k,{#1}}(\tilt)}
\newcommand{\fading}[1]{\Omega_{k,#1}}
\newcommand{\mfactor}[1]{m_{#1}}
\newcommand{\losP}[2]{p_{#1\hspace{-0.6mm}}\left( {#2} \right) }

\newcommand{\vGain}{G\left(\hd ,\tilt \right) }

\newcommand{\servingBS}{\mathbf{x}_\tau}

\newcommand{\lambdaBS}{\lambda_\text{B}}

\newcommand{\lambdaUp}{\lambda^\text{ES}_\text{B,A}}
\newcommand{\lambdaDown}{\lambda^\text{ES}_\text{B,G}}

\newcommand{\lambdag}{\lambda_{\text{G}}}
\newcommand{\lambdaa}{\lambda_{\text{A}}}
\newcommand{\lambdaI}{ \lambda_{\text{I}}	}
\newcommand{\lambdaInt}[2]{\lambda^{#1}_{\text{I},#2}}

\newcommand{\ccdf}[2]{\bar{{F}}^{s}_{#1}(#2)}

\newcommand{\pdf}[2]{f^{s}_{#1}(#2)}
\newcommand{\rvX}{X_{k}^{vj}}
\newcommand{\rvr}{\shd^{vj}}
\newcommand{\rvlambda}{\lambda_{\text{B},k}^{s}}

\newcommand{\PoutIS}
{\hat{\mathcal{P}}_{\text{no}}^{\text{IS}}\left( \theta_{\text{t,0}} \right) }
\newcommand{\PoutES}{\hat{\mathcal{P}}_{\text{no}}^{\text{ES}}\left( \tiltgES, \tiltaES \right) }
\newcommand{\PoutS}{\hat{\mathcal{P}}_{\text{no}}^{s}\left( \tiltg,\tilta \right) }

\newcommand{\PoutISop}{\hat{\mathcal{P}}_{\text{no}}^{\text{IS}}\left( \optiltIS \right) }
\newcommand{\PoutESop}{\hat{\mathcal{P}}_{\text{no}}^{\text{ES}}\left( \optiltg, \optilta \right) }

\newcommand{\pout}[3]{\mathcal{P}^{#1}_{\text{o},j}(#2,#3) }
\newcommand{\npout}[3]{\hat{\mathcal{P}}^{#1}_{\text{o},j}(#2,#3) }
\newcommand{\laplace}[1]{\mathcal{L}_{#1}(z)}
\newcommand{\ap}{\mathcal{A}_{vj}^{a}}

\newcommand{\prob}[1]{\mathbb{P}\left[#1\right]}

\newcommand{\arct}[1]{\arctan\left( #1\right)}

\newcommand{\bi}{\begin{itemize}}
\newcommand{\ei}{\end{itemize}}

\newcounter{eqncnt}

\newcounter{eqnback}

\begin{document}

\newcommand{\paperTitle}
{
Non-Terrestrial Networks for UAVs:\\ Base Station Service Provisioning Schemes \\ with Antenna Tilt
}
%


\title{\paperTitle}

\author{
	\vspace{0.2cm}
        Seongjun~Kim, Minsu~Kim, Jong~Yeol~Ryu, \textit{Member, IEEE},
        \\
          Jemin~Lee, \textit{Member, IEEE}, and Tony~Q.~S.~Quek, \textit{Fellow, IEEE}
    \thanks{
    	The material in this article was presented, in part, at the Global Communications
    	Conference, Taipei, Taiwan, Dec. 2020 \cite{KimKimLee:20}.
    }
        \thanks{
        	S.\ Kim and M.\ Kim are with the Department of Information and Communication Engineering, Daegu Gyeongbuk Institute of Science and Technology (DGIST), Daegu 42988, South Korea 
        	(e-mail: kseongjun@dgist.ac.kr; ads5577@dgist.ac.kr). 
    	
    	J.\ Y.\ Ryu is with the Department of Information and Communication Engineering, Gyeongsang National University, Tongyeong 53064, South Korea (e-mail: jongyeol\_ryu@gnu.ac.kr)
    	
    	J.\ Lee is with the Department of Electrical and Computer Engineering, Sungkyunkwan University (SKKU), Suwon 16419, South Korea  (e-mail: jemin.lee@skku.edu)
  
  T.\ Q.\ S.\ Quek is with the Information Systems Technology and Design Pillar, the Singapore University of Technology and Design, Singapore 487372 (e-mail: tonyquek@sutd.edu.sg).
    
        }
%
%
%
}

\maketitle 
%

%

%
\setcounter{page}{1}
\acresetall
\begin{abstract}
By focusing on \ac{UAV} communications in \acp{NTN}, this paper provides a guideline on the appropriate \ac{BS} service provisioning scheme with considering the antenna tilt angle of \ac{BS}.
Specifically, two service provisioning schemes are considered including the \ac{IS-BS} scheme, which makes \acp{BS} serve both \acp{GU} and \acp{AU} (i.e., \acp{UAV}) simultaneously, and the \ac{ES-BS} scheme, which has \acp{BS} for \acp{GU} and \acp{BS} for \acp{AU}.
By considering the antenna tilt angle-based channel gain, we derive the network outage probability for both \ac{IS-BS} and \ac{ES-BS} schemes, and show the existence of the optimal tilt angle that minimizes the network outage probability after analyzing the conflict impact of the antenna tilt angle.
We also analyze the impact of various network parameters, including the ratio of \acp{GU} to total users and densities of total and interfering \acp{BS}, on the network outage probability.
Finally, we analytically and numerically show in which environments each service provisioning scheme can be superior to the other one.
\end{abstract}


\acresetall


\section{Introduction}

Due to an increasing demand for novel and high-quality mobile services, 
it becomes more difficult to provide reliable communications by the existing terrestrial networks only, up to the level required by future mobile services.
To address these issues, \acp{NTN} have been considered as a promising solution to complement terrestrial networks by providing ubiquitous and global connectivity \cite{RinFedHel:20,GioZor:21}.
Conventional 2D ground space in terrestrial networks is now expanded to 3D aerial space in \acp{NTN} with supporting communications for \acp{UAV}, \ac{HAPS}, and satellites \cite{XinSteSeb:21}.
Among them, \ac{UAV} communications have been in the spotlight 
because \acp{UAV} have more flexible mobility and can locate closer to ground users and \acp{BS} in terrestrial networks, compared to \ac{HAPS} and satellites.
Therefore, many applications and services based on \ac{UAV} communications have appeared
such as working as a relay in  hotspot and a data collector in large-scale networks \cite{YonRuiTen:16b,NasMotMil:17,SamEvsRah:16}.
%
%
However, the integration of \acp{UAV} into existing terrestrial networks brings a lot of challenges such as resource and interference management since \ac{UAV} communications usually use the frequency band as well as \acp{BS} of terrestrial networks. 
%


In this context, many works have been presented for reliable \ac{UAV} communications.
At the beginning of studies, the wireless channel modeling of \ac{UAV} networks has been studied in \cite{AziYunNan:18,AkrKar:18,AlhKanLar:14,ChoLiuLee:18}, which is different from that of terrestrial networks.
Specifically, according to the height of the \ac{UAV}, the distance-dependent path loss model for the cellular-to-\ac{UAV} channel and the \ac{LoS} probability between the \ac{UAV} and the ground device were modeled in \cite{AziYunNan:18,AkrKar:18} and \cite{AlhKanLar:14,ChoLiuLee:18}, respectively.
Based on the wireless channel modeling of \ac{UAV} networks, the works in \cite{AzaRosChe:18,PanDon:17,HaiShuYon:18,MohWalMer:16,JiaYonRui:18,ShoYonRui:19,KimLee:19,KimLeeQue:20} studied to present the optimal location of \acp{UAV} for various environments and applications.
The deployment and the power allocation for the \ac{UAV} jointly optimized to minimize the outage probability in \cite{AzaRosChe:18,PanDon:17}.
The height of the \ac{UAV} and the antenna beamwidth jointly optimized to maximize the data rate \cite{HaiShuYon:18} and the coverage probability \cite{MohWalMer:16}.
The joint optimization of the \ac{UAV} trajectory and the spectrum allocation were considered  to maximize the throughput \cite{JiaYonRui:18} and  minimize the mission completion time \cite{ShoYonRui:19}.
The outage probability was presented by considering the effect of the \ac{UAV} height and the channel environment in \cite{KimLee:19}. 

In \cite{KimLeeQue:20}, multi-layer aerial networks have been considered and designed optimally to maximize the successful transmission probability and the area spectral efficiency.
However, the works in \cite{PanDon:17,HaiShuYon:18,JiaYonRui:18,ShoYonRui:19} made a strict assumption that \ac{UAV}-to-ground communications channels are dominated by \ac{LoS} links only without considering the location-dependent probability of having \ac{LoS} links.
Furthermore,  all of those  aforementioned works did not consider a \ac{BS} antenna tilt angle, which significantly affects the communication performance between the ground \ac{BS} and the \ac{UAV}.
Especially, the antenna tilt angle of the ground \ac{BS} has been conventionally designed for ground devices only, so the \ac{UAV} can actually receive the signal from these \acp{BS} with considerably small power \cite{YonJiaRui:19,XinVijSiv:18}.

To overcome these issues, the efficient design of the \ac{BS} antenna tilt angle for \ac{UAV} communications has been considered in recent works \cite{GalKibDas:18,AzaRosPol:17,AzaRosPol:18,RamWalNic:19,XiaYon:19,RamWalBor:20,RamWalNic:20}.
The vertical antenna gain was considered for analyzing the successful transmission probability of \ac{UAV} communications in \cite{GalKibDas:18}. 
The \ac{BS} antenna tilt angle was optimized to maximize the coverage probability according to the heights of the \ac{UAV} and the \ac{BS} in \cite{AzaRosPol:17,AzaRosPol:18}, and also to maximize the successful content delivery probability in massive \ac{MIMO} systems in \cite{RamWalNic:19}.
The \ac{BS} association probability and \acp{SINR} were studied for two different association policies such as nearest-distance based and maximum-power based associations by considering the antenna gain, determined by the tilt angle in \cite{XiaYon:19}.
The handover rate as well as the coverage probability were analyzed by considering the practical antenna configuration \cite{RamWalBor:20} and also for the \ac{CoMP} transmission \cite{RamWalNic:20}.

However, the aforementioned works considered limited scenarios and parameters of \ac{UAV} networks in the design of the antenna tilt angle.  
For instance, in \cite{GalKibDas:18,RamWalBor:20,RamWalNic:20}, a simple \ac{UAV} network, where \acp{GU} do not exist, was considered in spite of using ground \acp{BS}. 
In \cite{GalKibDas:18,AzaRosPol:17,AzaRosPol:18,XiaYon:19,RamWalNic:19,RamWalBor:20,RamWalNic:20}, they considered either the down tilt angle  or the up tilt angle although both should be considered to support \acp{AU} together with \acp{GU}. 
In \cite{AzaRosPol:17,AzaRosPol:18,RamWalNic:19,RamWalNic:20}, a simple constant power gain model was used for the antenna main lobe although it can be changed according to the \ac{BS} antenna tilt angle as well as the elevation angle of the communications link \cite{3GPP:TR:36.814:V9.2.0}. 
%
%
%
%
%
Furthermore, only the \emph{\ac{IS-BS} scheme} that makes each \ac{BS} serves both \acp{GU} and \acp{AU} was explored as in \cite{GalKibDas:18,AzaRosPol:17,RamWalNic:19}.
However, the \emph{\ac{ES-BS} scheme} that makes \acp{BS} serve \acp{GU} or \acp{AU} exclusively might be a better scheme for certain \ac{UAV} network environments.
%
%
%
%
Therefore, in existing works, it failed to present or analyze the performance of \ac{UAV} communications with more realistic tilt angle-based antenna gains as well as existing both \acp{GU} and \acp{AU} in the networks.

Therefore, in this paper, we provide a framework to explore an appropriate \ac{BS} service provisioning scheme to support both \acp{GU} and \acp{AU} with considering the tile angle-based antenna gain. 
First, the network outage probabilities of the \emph{\ac{ES-BS} scheme} as well as the \emph{\ac{IS-BS} scheme} are analyzed. We then explore how the optimal antenna title angles of \acp{BS} that minimize the network outage probability are determined for different service provisioning schemes as well as different types of \acp{BS}. 
The impact of various network parameters such as the spatial densities of total \acp{BS} and interfering \acp{BS} on the performance of service provisioning schemes are also discussed.
%
%
%
%
%
The main contributions of this paper are summarized as follows.

\bi
\item
We newly derive the network outage probability for two \ac{BS} service provisioning schemes, i.e., \ac{IS-BS} and \ac{ES-BS} schemes, by considering the tilt angle-based antenna gain in both general environment (where the interference exists) and noise-limited environment.
\item 
We analytically show that changing the antenna tilt angle gives conflicting impacts on the network outage probability.
Specifically, as the absolute value of the tilt angle decreases, the service area with the main lobe becomes wider (i.e., positive impact), but the link distance between the serving \ac{BS}
and the user increases (i.e., negative impact).
From these results, we show that there exists the optimal \ac{BS} antenna tilt angle that minimizes the network outage probability.
%
%
\item
%
%
%
%
We show the impact of various network parameters on the optimal antenna tilt angle that minimizes the network outage probability including the \ac{BS} height, the \ac{UAV} height, the ratio of \ac{GU}, as well as densities of the total \acp{BS} and the interfering \acp{BS}.
For instance, we show that the optimal antenna tilt angle increases as the ratio of \acp{GU} increases, and the absolute value of the tilt angle increases, as the total \ac{BS} density increases.
\item
We also explore which service provisioning scheme can be better in terms of the network outage probability in various environments.
Specifically, in the noise-limited environment, we \emph{analytically} show the superiority of the \ac{ES-BS} scheme to the \ac{IS-BS} scheme for high \ac{BS} density regime.
In the general environment, we numerically show the superiority of the \ac{IS-BS} scheme for low interfering \ac{BS} density regime and that of the \ac{ES-BS} scheme for high interfering \ac{BS} density regime.
%
%
%
%
\ei

The rest of this paper is organized as follows.
In Section \ref{sec:models}, we represent the \ac{BS} service provisioning schemes to serve both types of users and describe the channel model and the \ac{BS} antenna power gain, which is affected by the \ac{BS} antenna tilt angle.
We then describe the \ac{BS} association rule.
In Section \ref{sec:analysis}, we derive the network outage probability for two service provisioning schemes in the general environment and the noise-limited environment, respectively.
In Section \ref{sec:numerical}, we evaluate the network outage probability according to the various network design parameters.
We then compare the communication performance of the \ac{IS-BS} scheme and that of the \ac{ES-BS} scheme for network parameters.
Finally, conclusions are presented in \ref{sec:conclusion}.

\begin{table}
	\caption{Notations used throughout the paper.} \label{table:notation}
	\vspace{-5mm}
	\begin{center}
		\rowcolors{2}
		{cyan!15!}{}
		\renewcommand{\arraystretch}{1.0}
		\begin{tabular}{ c  p{5.5cm} }
			\hline 
			{\bf Notation} & {\hspace{2.0cm}}{\bf Definition}
			\\
			\midrule
			\hline
			$i\in\left\lbrace \text{G}, \text{A} \right\rbrace$ & User type index for \acp{GU} ($i\hspace{-0.5mm}=\hspace{-0.5mm}\text{G}$) and \acp{AU} ($i\hspace{-0.5mm}=\hspace{-0.5mm}\text{A}$)
			\\ \addlinespace
			
			$l\in\left\lbrace \text{O}, \text{G}, \text{A} \right\rbrace$ &
			\ac{BS} type index for \ac{IS-BS} ($l=\text{O}$), GBS ($l=\text{G}$) and ABS ($l=\text{A}$)
			\\ \addlinespace
			$v\in \left\lbrace \text{L,N} \right\rbrace $ & Index for the \ac{LoS} environment $(v=\text{L})$ and the \ac{NLoS} environment $(v=\text{N})$           \\ \addlinespace
			$s\in\left\lbrace \text{IS},\text{ES}\right\rbrace $ & Index for the \ac{IS-BS} scheme $(s=\text{IS})$ and the \ac{ES-BS} scheme $(s=\text{ES}) $					  \\ \addlinespace 
			$\hd$ & Horizontal distance between the $k$th user and the \ac{BS} located at $\mathbf{x}$ \\ \addlinespace
			$\hBS$ / $\hi$ & Height of the \ac{BS} and the $k$th user 						 	  \\ \addlinespace	
			$\fading{\mathbf{x}}$ & 
			Channel fading between the $k$th user and the \ac{BS}   \\ \addlinespace
			$\tiltIS$ / $\tiltgES$, $\tiltaES$ & 
			Antenna tilt angle of the \ac{IS-BS} / Antenna tilt angles of GBS and ABS in the \ac{ES-BS} scheme.			  
			\\ \addlinespace
			$\tpower$ & Transmission power of the \ac{BS}     \\ \addlinespace 
			$\target$ & Target SINR/SNR						  \\ \addlinespace
			$\ld{\hd}$ & Path loss between the $k$th user and the \ac{BS}					  \\ \addlinespace
			$\losP{\text{L}}{\hd}$ / $\losP{\text{N}}{\hd}$ & \ac{LoS} and \ac{NLoS} probability for given $\hd$	  \\ \addlinespace
			$\vGain$ & Antenna power gain for given $\hd$ and $\tilt$       \\ \addlinespace
			$\loDistance{\text{G}}$ / $\loDistance{\text{A}}$ & 
			Lower bound of the horizontal distance that the user is served by main lobe for \acp{GU} and \acp{AU}	     
			\\ \addlinespace
			$\upDistance{\text{G}}$ / $\upDistance{\text{A}}$ &  			
			Upper bound of the horizontal distance that the user is served by main lobe for \acp{GU} and \acp{AU}		
			\\ \addlinespace
			$f_{\rvr}^{s,a}(r)$ & PDF of the horizontal distance between the $k$th user and the serving \ac{BS} for given scheme $\hspace{-0.2mm}s$ and association rule $\hspace{-0.2mm}a\hspace{-0.2mm}$ \\ \addlinespace
			$\rho_{i}$ & Ratio of $i$-type users to total users  \\ \addlinespace
			$\rho_{\text{B},i}$ & Ratio of \acp{BS} for $i$-type user to total \acp{BS} \\ \addlinespace
			$\lambdaBS$ / $\lambdaI$ & Densities of total \ac{BS} and total interfering \ac{BS} \\ \addlinespace
			$\lambda^s_{\text{B},i}$ & 
			\ac{BS} density for $i$-type user in scheme $s$
			\\ \addlinespace
			$\lambdaInt{\text{IS}}{\text{O}}$ / $\lambdaInt{\text{ES}}{\text{G}}$, $\lambdaInt{\text{ES}}{\text{A}}$ & 
			Density of interfering \ac{IS-BS} / Densities of interfering GBS and ABS in the \ac{ES-BS} scheme.				  
			\\ \addlinespace
			$\interference{\text{IS}}{\text{O}}$ / $\interference{\text{ES}}{\text{G}}$, $\interference{\text{ES}}{\text{A}}$ & 
			Interference from \acp{IS-BS} / Interference from GBSs and ABSs in the \ac{ES-BS} scheme.		
			\\ \addlinespace
			
			$\mathcal{P}_{\text{no}}^s \left( \tiltg, \tilta\right) $ & Network outage probability in the general environment for given $s$ \\ \addlinespace
			$\PoutS$ & Network outage probability in the noise-limited environment for given $s$ \\ \addlinespace
			$	\tilde{\mathcal{P}}^s_{\text{no}} \left(\tiltg,\tilta \right) $ & Network outage probability with simplified antenna gain model for given $s$.
			\\ \addlinespace
			\hline 
		\end{tabular}
	\end{center}\vspace{-0.63cm}
\end{table}%

\emph{Notation}: The notation used throughout the paper is listed in Table \ref{table:notation}.


\section{System Model}\label{sec:models}
In this section, we introduce the non-terrestrial network model by mainly focusing on \ac{UAV} networks.
Moreover, we describe the antenna power gain and the \ac{BS} association rules.
\subsection{Network Model} \label{subsec:netmodel}

We consider  a \ac{NTN} for \acp{UAV}, where \acp{BS}, ground users (GUs), and aerial users (AUs) (i.e., \acp{UAV}) are randomly distributed in the spatial domain.
%
%
%
%
The locations of users are modeled by \ac{HPPP} $\hpppi$ with density $\lambda_i$, where $i\in\left\lbrace \text{G},\text{A}\right\rbrace$ denotes the type of users, i.e., $i=\text{G}$ for \acp{GU} and $i=\text{A}$ for \acp{AU}.\footnote{
Note that we can also assume \acp{AU} are distributed according to \ac{MHCPP} with density $\lambdaa$ that considers the minimum safety horizontal distance, $d_\text{min}$, between any two \acp{AU} like the ones in \cite{ZhuZheFit:18,LyuWan:21}.
However, the performance analysis and the results of this work will be the same as only the density of \acp{AU} affects the performance, not the distribution as the downlink is considered.
}

The height of the $k$th user is $h_k$, where $h_k=\hGU$ for $k\in\mathcal{U}_\text{G}$ and $h_k=\hAU$ for $k\in\mathcal{U}_\text{A}$. Here $\mathcal{U}_\text{G}$ and $\mathcal{U}_\text{A}$ are the user index sets of \acp{GU} and \acp{AU}, respectively.
%
%
%
%
%
%
%
\begin{figure}[t!]
	\begin{center}   
		{ 
			\includegraphics[width=1.0\columnwidth]{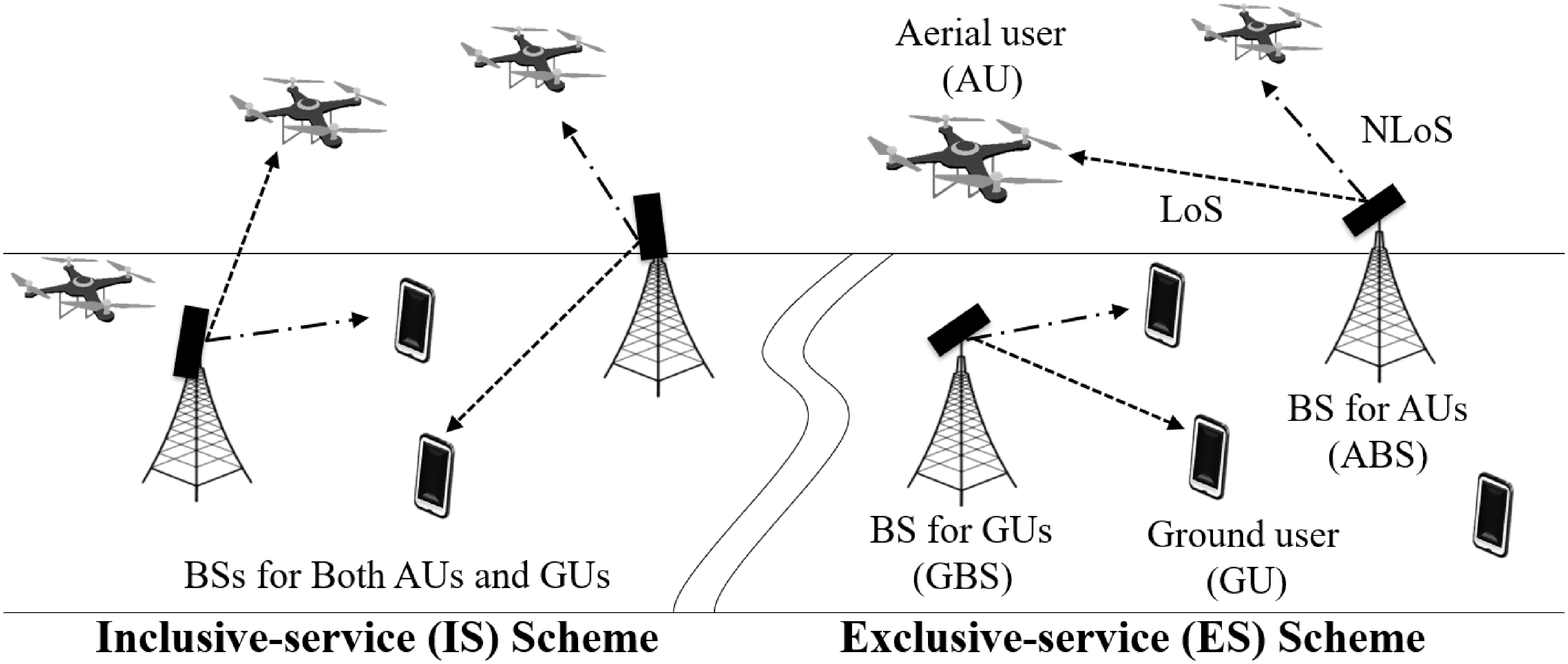}
		}
	\end{center}
	
	\caption{ Examples of \ac{NTN} for \acp{UAV} with randomly distributed \acp{BS}, \acp{GU}, and \acp{AU}.
	}
	\label{fig:NetworkModel}
	\vspace{-1.5mm}
\end{figure}

%
%
%

In this paper, as shown in Figure \ref{fig:NetworkModel}, we consider the two types of \ac{BS} service provisioning schemes as follows.
\bi
\item
\emph{Inclusive-service BS (\ac{IS-BS}) scheme}: 
In this scheme, \acp{BS} serve both \acp{GU} and \acp{AU} simultaneously.
Hence, the antenna tilt angle of the \ac{BS} has to be designed to serve both \acp{GU} and \acp{AU} efficiently.
The locations of \acp{BS} are modeled by \ac{HPPP} $\hpppBS{\text{O}}{}$ with density $\lambdaBS$.
Since there is only one type of \ac{BS}, the \ac{BS} density for \acp{GU}, $\lambda^\text{IS}_{\text{B,G}}$, and the \ac{BS} density for \acp{AU}, $\lambda^\text{IS}_{\text{B,A}}$, are the same as the total \ac{BS} density (i.e., $\lambda^\text{IS}_{\text{B,G}}=\lambda^\text{IS}_{\text{B,A}}=\lambdaBS$).	
Note that this scheme is the one, generally used in prior works such as \cite{GalKibDas:18,AzaRosPol:17,RamWalNic:19}.
%
%
%
\item
\emph{Exclusive-service \ac{BS} (\ac{ES-BS}) scheme}: 
In this scheme, \acp{BS} are divided into two groups: 1) a \ac{BS} for \acp{GU} (\ac{GBS}) and 2) a \ac{BS} for \acp{AU} (\ac{ABS}).
The \acp{GBS} and \acp{ABS} exclusively serve \acp{GU} and \acp{AU}, respectively.
Therefore, the antenna tilt angles of \acp{GBS} and \acp{ABS} need to be designed respectively to serve aimed users efficiently.
We assume the distributions of \acp{GBS} and \acp{ABS} also follow \acp{HPPP}, $\hpppBS{\text{G}}{}$ and $\hpppBS{\text{A}}{}$, with densities $\lambdaDown=\bsratio\lambdaBS$ and $\lambdaUp=(1-\bsratio)\lambdaBS$, respectively, where $\bsratio$ is the portion of \acp{GBS} among all \acp{BS}.
\ei
Regardless of \ac{BS} types, for all \acp{BS}, the antenna height is $\hBS$
and the transmission power is $\tpower$.

\subsection{Channel Model}
In \ac{UAV} communications, both \ac{LoS} and \ac{NLoS} environments can be considered for the links between a \ac{BS} and a \ac{GU} as well as between a \ac{BS} and an \ac{AU}. 
The probability of forming \ac{LoS} link between the \ac{BS} at $\mathbf{x}=(x_\text{B},y_\text{B},\hBS)$ and the $k$th user at $(x_k,y_k,h_k)$ is given by \cite{ZhiLaiGua:18}\footnote{
		The \ac{LoS} probability is also defined differently in \cite{AlhKanLar:14}. 
		However, it is determined by the elevation angle between the transmitter and the receiver, not by the link distance.
	}
%
%
%
%
%
%
%
%
%
%
\begin{align}
\losP{\text{L}}{\hd}=& \label{eq:LoS_probability}
\left\{ \hspace{-0.5mm}
	1-\frac{\sqrt{2\pi}\xi}{\left| h_k \hspace{-0.5mm} - \hspace{-0.5mm} \hBS\right|} 
		\hspace{-0.5mm}	
		\left| 
			Q\hspace{-0.5mm}\left(\hspace{-0.5mm}	 \frac{\hi}{\xi}  \hspace{-0.5mm}	\right) 
			- 
			Q\hspace{-0.5mm}\left( \frac{\hBS}{\xi} \hspace{-0.5mm}  \right) \hspace{-0.5mm}
		\right| 
		\hspace{-0.5mm}	
\right\}^
{\hd \sqrt{\mu\nu} }  \hspace{-2mm},
\end{align}
where $Q(x)\hspace{-1mm}=\hspace{-1mm}\int_{x}^{\infty}\! \frac{1} {\sqrt{2\pi}} \exp (- \frac{t^2} {2}) \,dt$  is the Q-function and the horizontal distance between the \ac{BS} and the $k$th user is given by $\hd \hspace{-1mm} = \hspace{-1mm} \sqrt{\left(x_k\hspace{-1mm}-\hspace{-1mm}x_\text{B} \right)^2\hspace{-1mm}+\hspace{-1mm} \left(y_k-y_\text{B} \right)^2}$.
Here, $\mu$, $\nu$ and $\xi$ are environment parameters determined by the density and the height of obstacles. 
Since the \ac{NLoS} environment is a complementary event of the \ac{LoS} environment,
the \ac{NLoS} probability between the \ac{BS} and the $k$th user is given by $\losP{\text{N}}{\hd} = 1-\losP{\text{L}}{\hd}$.

Based on the \ac{LoS} probability, we consider different path loss exponents and channel fading models for \ac{LoS} and \ac{NLoS} links.
The path loss exponent for \ac{LoS} and \ac{NLoS} links are denoted by  $\alpha_\text{L}$ and $\alpha_\text{N}$, respectively.
The channel fading is modeled by Nakagami-$m$ fading, so the distribution of the channel gain is given by	
%
%
%
%
%
\begin{align} \label{eq:gamma}
f_{\Omega_v}(x)
=
\frac{\mfactor{v}^{\mfactor{v}}}{\Gamma(\mfactor{v})}
		\hspace{0.3mm} x^{\mfactor{v}-1} \hspace{0.3mm}
			\exp\left(\hspace{0.3mm} {-\mfactor{v}x}\hspace{0.3mm}\right),
			\hspace{3mm}
			x>0
			, 
\end{align}
where $\Gamma(x)=\int_{0}^{\infty}t^{x-1}e^{-t}dt$ and $v\in\{\text{L},\text{N}\}$ is the channel environment, i.e., $v=\text{L}$ for \ac{LoS} links and $v=\text{N}$ for \ac{NLoS} links.
In addition, we assume that $\mfactor{\text{L}}>1$, and $\mfactor{\text{N}}=1$, which means Rayleigh fading, i.e., $\Omega_\text{N} \sim \text{exp}(1)$.
From \eqref{eq:gamma}, we denote the channel fading between the $k$th user and the \ac{BS} as
%
%
%
\begin{align}
	\fading{\mathbf{x}}
	=
\begin{dcases}
	\Omega_\text{L},
	& \text{with probability} \ \losP{\text{L}}{\hd}
	\\
	\Omega_\text{N}, 
	&\text{otherwise}
\end{dcases}. 
\end{align}
%
%
%
%

\begin{figure}[t!]
	\begin{center}   
		{ 
			\includegraphics[width=1.00\columnwidth]{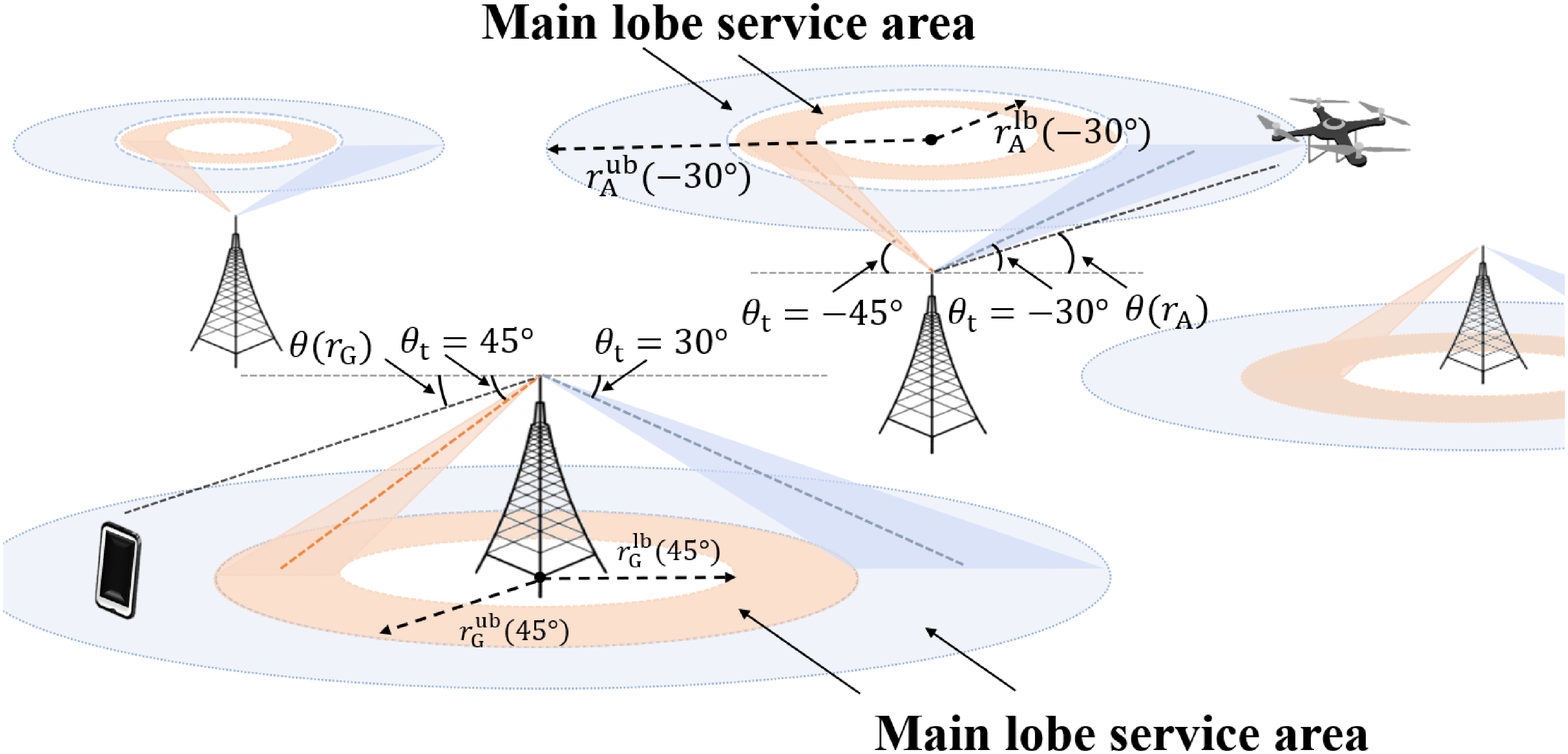}
		}
	\end{center}
	\caption{ 
		Examples of antenna radiation patterns for different antenna tilt angles. 
	}
	\label{fig:AntennaConfiguration}
\end{figure}

\subsection{Vertical Antenna Gain} \label{subsection:antenna_gain}
The antenna power gain of the \ac{BS} is determined by two types of power gains: horizontal and vertical directional antenna gains.
We consider an omnidirectional antenna in the horizontal direction, so the horizontal directional antenna gain remains constant regardless of the direction of the antenna.
Therefore, we assume the horizontal directional antenna gain is equal to a unit gain \cite{HerZaiMcl:15}.
In this paper, we focus on the design of the vertical antenna tilt angle for \acp{AU} as well as \acp{GU},
and we consider the directional antenna in the vertical direction. 
As shown in Fig. \ref{fig:AntennaConfiguration}, the vertical directional antenna gain is determined by the vertical antenna tilt angle, $-90^\circ \hspace{-0.5mm}<\hspace{-0.5mm} \tilt \hspace{-0.5mm}<\hspace{-0.5mm} 90^\circ$, which is the angle tilted upward or downward relative to the horizontal plane.\footnote{Note that there are two types of tilting methods\cite{AhnLee:14}: mechanical tilting and electrical tilting. 
The mechanical tilting rotates the antenna of the \ac{BS} physically.
On the other hand, the electrical tilting applies an overall phase shift to all antenna elements in the array. 
In this paper, we consider the electrical tilting method to analyze the communication performance mathematically.}
Here, we define that the \ac{BS} antenna tilt angle is a negative value when the \ac{BS} antenna tilt angle is up-tilted, i.e., tilting upwards with respect to the horizontal plane of the BS antenna. 
On the other hand, the \ac{BS} antenna tilt angle is defined as a positive value when the \ac{BS} antenna tilt angle is down-tilted, i.e., tilting downwards with respect to the horizontal plane of the BS antenna.
Based on the \ac{3GPP} specification \cite{3GPP:TR:36.814:V9.2.0}, for given $\hd$, the \ac{BS} antenna power gain $G\hspace{-0.3mm}(\hspace{-0.3mm}\hd,\tilt\hspace{-0.3mm})$  can be represented by a function of the tilt angle as 
%
%
%
\begin{align} \label{eq:AG_3GPP}
\vGain
=
10^{-\min
		\left(12 
			\left(
					\frac{\eangle+\tilt}{\angledb} 
			\right)^2, \minpower
		\right)/10
	},
\end{align}  
where $\angledb=10^\circ$ is the 3dB beamwidth and $\minpower$ is the minimum power leaking to the side lobe besides the main lobe, which is commonly 20dB. 
In \eqref{eq:AG_3GPP}, $\eangle$ is the elevation angle between the \ac{BS} antenna and the $k$th user, which is given by
%
%
%
%
\begin{align}
\eangle=\frac{180}{\pi}\arct{\frac{h_k-\hBS}{\hd}},
\end{align}
where $h_k-\hBS$ is the height difference between the \ac{BS} and the $k$th user.
In this work, without loss of generality, we assume that the height of \acp{AU} is higher than that of \acp{BS} (i.e., $\hAU-\hBS>0$), while the height of \acp{GU} is lower than that of \acp{BS} (i.e., $\hGU-\hBS<0$).
From \eqref{eq:AG_3GPP}, for given $\tilt$, the user can be served with the main lobe when $12 \! \left( \! \frac{\eangle+\tilt}{\angledb} \! \right)^{\!\!2}  \geq \minpower$.
Here, we define the boundary of horizontal distance between a \ac{BS} and the $k$th user that the user is served by the main lobe as $\loDistance{k}\leq\hd\leq\upDistance{k}$, where $\loDistance{k}$ and $\upDistance{k}$ are the lower and upper boundaries.
Since all \acp{GU} and all \acp{AU} have the same height, $\hGU$ and $\hAU$, respectively, the boundaries are determined by the user types not user's specific location, i.e. $\upDistance{k}=\upDistance{i}$ and $\loDistance{k}=\loDistance{i}$ for $k\in\mathcal{U}_i$, and given as follows
%
%
%
%
%
%
%
%
%
%
\begin{align} 
&\loDistance{\text{G}}= \label{eq:lowerbound_g}
\begin{dcases}
	\frac{\hGU-\hBS}
	{\tan\left\{\frac{\pi}{180}\left(- \tilt -\angleth \right)\right\}},
	& \tilt > -\angleth
\\
	\quad \quad \quad \,\,\,\, \infty, & \text{otherwise}
\end{dcases},
\\
&\upDistance{\text{G}}=\label{eq:upperbound_g}
\begin{dcases}
	\frac{\hGU-\hBS}
	{\tan\left\{\frac{\pi}{180}\left(-\tilt + \angleth \right)\right\}},
	&  \tilt > \angleth
\\
	\quad \quad \quad \,\,\,\,\,\, \infty, & \text{otherwise}
\end{dcases},
\\
&\loDistance{\text{A}}=\label{eq:lowerbound_a}
\begin{dcases}
	\frac{\hAU-\hBS}
	{\tan\left\{\frac{\pi}{180}\left(- \tilt + \angleth \right)\right\}},
	&  \tilt < \angleth
\\
	\quad \quad \quad \,\,\,\,\,\, \infty, & \text{otherwise}
\end{dcases}
\\
&\upDistance{\text{A}}=\label{eq:upperbound_a}
\begin{dcases}
	\frac{\hAU-\hBS}
	{\tan\left\{\frac{\pi}{180}\left(- \tilt - \angleth \right)\right\}},
&  \tilt < -\angleth
\\
	\quad \quad \quad \,\,\,\,\,\, \infty, & \text{otherwise}
\end{dcases},
\end{align}
where $\angleth=\angledb\sqrt{ \minpower/12}$.
In \eqref{eq:lowerbound_g}-\eqref{eq:upperbound_a}, the boundaries $\upDistance{i}$ and $\loDistance{i}$ are defined to be positive when $\tilt$ satisfies each conditions.
For the convenience of analysis, we rewrite $\vGain$ in \eqref{eq:AG_3GPP} according to the boundaries in \eqref{eq:lowerbound_g}-\eqref{eq:upperbound_a} as
%
%
%
%
%
\begin{align}	\label{eq:AG}
\vGain=
\begin{dcases}
	G_1(\hd,\tilt),
	& \bound{1} < \hd < \bound{2}
\\
	G_2(\hd,\tilt), 
	&\bound{2}\leq \hd  \leq \bound{3}
\\
	G_3(\hd,\tilt),
	 & \bound{3} < \hd < \bound{4}
\end{dcases},
\end{align}
where $\bound{1}=0, \bound{2}=\loDistance{k}, \bound{3}=\upDistance{k}$,  and $\bound{4}=\infty$.
In \eqref{eq:AG}, $G_1(\hd, \tilt)=G_3(\hd, \tilt)=10^{-\minpower/10}$ is the antenna side lobe gain and $G_2(\hd, \tilt)$ is the antenna main lobe gain, which is given by
%
%
%
%
%
\begin{align}	\label{eq:AG2}
	G_2(\hd,\tilt)=10^{-1.2
		\left\lbrace 
			\frac{\eangle+\tilt}{\angledb}	
			\right\rbrace^2}.
\end{align}
From \eqref{eq:AG2}, we can see that $G_2(\hd, \tilt)$ is an increasing function of $\tilt$ for $-\angleth< \tilt \leq -\eangle$, and $G_2(\hd, \tilt)$ is a decreasing function of $\tilt$ for $-\eangle \leq \tilt < \angleth$.
This is because as the antenna tilt angle $\tilt$ approaches the elevation angle between the \ac{BS} and the user, the effect of the main lobe becomes dominant and it is maximized when the antenna tilt angle is equal to the elevation angle (i.e., $\tilt =-\eangle$).

\subsection{\ac{BS} Association Rule}

\setcounter{eqnback}{\value{equation}}
\setcounter{equation}{15}
\begin{figure*}[t!]
	\begin{align} \label{eq:prob_str}
		\prob{\rvX \leq r, \mathbf{x}_\tau \in \hpppBS{l}{vj} \left| \ a=\text{sa} \right. }
		\hspace{-1mm}
		\overset{\underset{\mathrm{(a)}}{}}{=}
		\hspace{-1mm}
		\int_{b_{k,j}(\tilt)}^{r}
		\hspace{-1mm}
		\pdf{\rvX}{x}
		\! \! \! \!\!\!\!\!\!\!\!\!\!
		\prod_{
			\substack{
				j_o\in \mathcal{J},
				v_o\in\left\lbrace \text{L},\text{N}\right\rbrace
				,
				\\
				(j_o,v_o) \neq (j,v) 
			}
		}^{}
		\!\!\!\!\!\!\!
		\prob{
			G_j(x,\tilt)	\hspace{-1mm}
			\left( x^2 +h_k^2\right) ^{\!\!-\!\frac{\alpha_v}{2}}
			\hspace{-0.5mm}
			\geq
			\hspace{-0.5mm}
			G_{j_o}\hspace{-1mm}\left( X_k^{v_oj_o},\tilt\right)\hspace{-1.5mm}
			\left( \hspace{-1mm} 
			(X_k^{v_oj_o})^2 +h_k^{2}\right)^{\!\!-\!\frac{\alpha_{v_o}}{2}} 
		}\hspace{-1mm} dx.
	\end{align}
	\setcounter{eqncnt}{\value{equation}}
	\centering \rule[0pt]{18cm}{0.3pt}
	\vspace{-3.0mm}
\end{figure*}
\setcounter{equation}{\value{eqnback}}

\label{subsec:association_rule}
In conventional networks, the \ac{BS} association is determined by the mean channel fading gain and the distance-dependent path loss, considering the \ac{LoS} probability \cite{MohHal:18}. 
However,  in more realistic \ac{UAV} networks, the antenna gain $\vGain$ affected by the horizontal distance between the serving BS and the $k$th user should also be considered in the \ac{BS} association.

To analyze \ac{BS} association rules, we first denote \acp{BS} which belong to $\hpppBS{l}{}, l\in{\text{\{O,G,A\}}}$, forming \ac{LoS} and \ac{NLoS} links as $\hpppBS{l}{\text{L}}$ and $\hpppBS{l}{\text{N}}$, respectively.
We then divide each of $\hpppBS{l}{\text{L}}$ and $\hpppBS{l}{\text{N}}$ into three groups  according to the \ac{BS} antenna power gain $\vGain$ in \eqref{eq:AG} as
\begin{align}  \label{eq:layer}
&\hpppBS{l}{vj}	=
\begin{dcases}
\hpppBS{l}{v1},
&  \bound{1} < \hd < \bound{2}
\\
\hpppBS{l}{v2}, 
& \bound{2}  \leq \hd \leq \bound{3}
\\
\hpppBS{l}{v3}, 
& \bound{3}  < \hd < \bound{4}
\end{dcases},~v\in\{\text{L},\text{N}\},
\end{align}
where $j\hspace{-1mm}\in\hspace{-1mm}\mathcal{J}$ is the index of \ac{BS} groups which is determined by $\hd$, and  $\mathcal{J}\hspace{-1.5mm}=\hspace{-1.5mm}\{\hspace{-0.5mm}1,2,3\hspace{-0.5mm}\}$.
Note that from \eqref{eq:AG} and \eqref{eq:layer}, we know that \acp{BS} in $\hpppBS{l}{v1}$ or $\hpppBS{l}{v3}$ transmit with the antenna side lobe gain, and \acp{BS} in $\hpppBS{l}{v2}$ transmit with the antenna main lobe gain. 
First of all, we examine the distribution of the distance between the user and the \ac{BS} in $\hpppBS{l}{vj}$ . 
The horizontal distance to the nearest \ac{BS} among the \acp{BS} in $\hpppBS{l}{vj}$ is denoted by $\rvX$.
Here, depending on the \ac{LoS} probability, the density function of $\hpppBS{l}{vj}$ is given by $2\pi \rvlambda t p_{v}(t)$. 
Therefore, for \acp{BS} in $\hpppBS{l}{vj}$, the \ac{CCDF} of $\rvX$ can be obtained as
\begin{align}
\ccdf{\rvX}{x} \label{eq:ccdf}
\hspace{-0.7mm}=\hspace{-0.7mm}
\mathbb{P}\hspace{-0.5mm}\left[ \hspace{-0.3mm}
				\rvX \!\geq \!x
				\hspace{-0.3mm}
\right] 
\hspace{-0.7mm}
\overset{\underset{\mathrm{(a)}}{}}{=} \hspace{-0.7mm}
\exp  \hspace{-0.7mm}
\left\{\hspace{-0.7mm}
- 2 \pi \rvlambda \hspace{-1.5mm}
\int_{\bound{j}}^{u_{k,j}(x,\tilt)} t p_{v}(t) dt
\!\right\}\hspace{-1mm},          
\end{align}
where (a) is from the void probability\cite{BacBla:09b} and
$u_{k,j}(x,\tilt)$ is given as  $u_{k,1}(x,\tilt)=\text{min}(x,\bound{j+1})$ if $j=1$, $u_{k,j}(x,\tilt)=\text{max}(x,\bound{j})$, otherwise.
$\lambda_{\text{B},k}^{s}$ is the density of \acp{BS} that can serve the $k$th user, i.e., $\lambda_{\text{B},k}^{s}=\lambda_{\text{B},i}^{s}$ when $k\in\mathcal{U}_i$.
Here, $s\in\left\lbrace \text{IS},\text{ES}\right\rbrace $ is the index of the \ac{BS} service provisioning scheme.
By differentiating \eqref{eq:ccdf}, we can obtain the \ac{PDF} of $\rvX$ as
\begin{align}
%
\pdf{\rvX}{x} 
&= \label{eq:pdf}
2 \pi \rvlambda x p_v(x) \exp  \left\{
- 2 \pi \rvlambda 
\int_{\bound{j}}^{x} t p_{v}(t) dt
\right\},
\end{align}
where $\pdf{\rvX}{x}=0$ if $x \leq \bound{j}$.

We denote $a\in \{{\text{na},\text{sa}}\}$ as the index of the \ac{BS} association criterion. Here, $a=\text{na}$ and $a=\text{sa}$ indicate the nearest \ac{BS} association rule and the strongest \ac{BS} association rule, respectively.

\subsubsection{Nearest \ac{BS} Association Rule}
In the nearest \ac{BS} association rule ($a=\text{na}$), the horizontal distance between the $k$th user and the serving \ac{BS} is smallest.
Therefore, the probability that the serving \ac{BS} exists in $\hpppBS{l}{vj}$, and the horizontal distance between the serving \ac{BS} and the $k$th user is smaller than $r$ is given by
\begin{align} \label{eq:prob_near}
&\prob{\rvX \leq r, \servingBS \in \hpppBS{l}{vj} \left| \ a=\text{na}\right. }
\nonumber
\\
&\overset{\underset{\mathrm{(a)}}{}}{=}
\int_{b_{k,j}(\tilt)}^{r}
\pdf{\rvX}{x}
\! \! \! \!\!\!
\prod_{
\substack{
	j_o\in \mathcal{J},
	v_o\in\left\lbrace \text{L},\text{N}\right\rbrace
	,
	\\
	(j_o,v_o) \neq (j,v) 
}
}^{}
\!\!\!\!\!\!\!
\prob{ x \leq X_k^{v_oj_o}
} dx.
\nonumber
\\
&=
\int_{b_{k,j}(\tilt)}^{r}
\pdf{\rvX}{x}
\! \! \! \!\!\!
\prod_{
\substack{
	j_o\in \mathcal{J},
	v_o\in\left\lbrace \text{L},\text{N}\right\rbrace
	,
	\\
	(j_o,v_o) \neq (j,v) 
}
}^{}
\!\!\!\!\!\!\!
\ccdf{X_k^{v_o j_o}}{x}
 dx,
\end{align}
where $\servingBS$ denotes the location of the serving \ac{BS} and (a) is from the fact that for given $j$ and $v$, the horizontal distance between the serving \ac{BS} and the user is shorter than all other candidates.

\subsubsection{Strongest \ac{BS} Association Rule}
In the strongest \ac{BS} association rule ($a=\text{s}$), the main link has the strongest average received power. 
The probability that the serving \ac{BS} exists in $\hpppBS{l}{vj}$ and the horizontal distance between the serving \ac{BS} and the $k$th user is smaller than $r$ is given in \eqref{eq:prob_str}, shown at the top of this page.
In \eqref{eq:prob_str}, (a) is from the fact that for given $j$ and $v$, the average power of the serving \ac{BS} should be greater than all other candidates.

From \eqref{eq:prob_near} and \eqref{eq:prob_str}, given $a\in\{\text{na},\text{sa}\}$, we can obtain the association probability $\ap$ as
\setcounter{equation}{16}
\begin{align} \label{eq:association_prob}
\ap
=
\prob{\rvX \leq \bound{j+1}, \servingBS \in \hpppBS{l}{vj} \left| \ a\right. }.
\end{align}
Therefore, when the $k$th user is associated with a BS in $j$-group under the channel environment $v$, the \ac{CDF} of the horizontal distance between the \ac{BS} and the user, $\rvr$, is given by 
\begin{align} \label{eq:cdf_association}
F^{s,a}_{\rvr}(r)
=
\prob{\rvX \leq r, \mathbf{x}_\tau \in \hpppBS{l}{vj} \left| \ a \right. }/\ap.
\end{align} 
By differentiating \eqref{eq:cdf_association}, we can obtain the \ac{PDF} of $\rvr$
\begin{align} \label{eq:link_pdf}
f^{s,a}_{\rvr \hspace{-1.5mm}}(r)
\hspace{-1mm}
=
\hspace{-1mm}
\begin{dcases}
\frac{2 \pi \rvlambda x p_{v\hspace{-0.5mm}}(x)}{\mathcal{A}_{vj}^{\text{na}}}
\!
 \exp  \hspace{-0.7mm}
 \left\{
 \hspace{-1.2mm} 
- 2 \pi \rvlambda \hspace{-1.5mm}
\int_{\bound{j}}^{x} \hspace{-7mm} t p_{v\hspace{-0.5mm}}(t) dt \hspace{-0.7mm}
\right\}\hspace{-1mm},
& \hspace{-2mm} a \hspace{-0.5mm}=\hspace{-0.5mm} \text{na}
\\
\frac{\partial}{\partial r}
F^{s,\text{sa}}_{\rvr}(r),
& \hspace{-2mm} a\hspace{-0.5mm}=\hspace{-0.5mm}\text{sa}
\end{dcases} \hspace{-2mm}.
\end{align}
Note that for the strongest association, $f^{s,\text{sa}}_{\rvr}(r)$ cannot be presented due to the complicated form of \eqref{eq:prob_str}.
However, in Section \ref{sec:numerical}, we show that the performance of the strongest association and that of the nearest association have similar  trends. This means we can use the analysis of the nearest association to design the case of the strongest association as well.


\section{Outage Probability Analysis}\label{sec:analysis}
In this section, for both \ac{IS-BS} and \ac{ES-BS} schemes, we derive the network outage probability in the presence of \acp{GU} and \acp{AU}. 
The outage probability is presented for two cases: the general environment in Section \ref{subsec:general} and the noise-limited environment in Section \ref{subsec:Noise} as a special case.

\subsection{General Environments}\label{subsec:general}
We assume that the available frequency resource is divided into $N$ sub-bands, and
the interfering \acp{BS} are the ones that use the same sub-band.
Hence, in the \ac{IS-BS} scheme, the distribution of the interfering \acp{BS} is modeled as a \ac{HPPP}  $\hpppI{\text{O}}$ with density $\lambdaInt{\text{IS}}{\text{O}}=\lambdaBS/N$ such as in \cite{HesAhmMoh:16}.
In the \ac{ES-BS} scheme, the interference from \acp{GBS} and \acp{ABS} needs to be defined differently as they use different tilt angles.
The distributions of interfering \acp{GBS} and \acp{ABS} are also modeled as \acp{HPPP}, $\hpppI{\text{G}}$ and $\hpppI{\text{A}}$, with densities $\lambdaInt{\text{ES}}{\text{G}}=\lambdaDown/N$ and $\lambdaInt{\text{ES}}{\text{A}}=\lambdaUp/N$, respectively.

For the case that a \ac{BS} communicates with the $k$th user, the \ac{SINR} at the user can be given by
%
%
%
%
%
%
\begin{align} \label{eq:SINR}
\gamma^s_v(\shd,\tilt) 
=
\frac
{\tpower \fading{v} \ld{\shd} G\left(\shd,\tilt \right) } 
{\interference{s}{}+\noise}
,
\end{align}
where $\ld{\shd}  \hspace{-0.8mm} = \hspace{-0.8mm}\left(\hspace{-0.5mm} \hd^2\hspace{-0.5mm}+\hspace{-0.5mm}\left(h_k\hspace{-1mm}-\hspace{-1mm}\hBS \right)^{2}\right)^{\hspace{-1.0mm}-\frac{\alpha_v}{2}}, v\in\{\text{L},\text{N}\}$, is the distance-dependent path loss between the $k$th user and the BS at $\mathbf{x}$ for \ac{LoS} and \ac{NLoS} links, and  $\noise$ is the noise power.

In \eqref{eq:SINR},  $\interference{\text{IS}}{}\hspace{-0.5mm}=\hspace{-0.5mm}\interference{\text{IS}}{\text{O}}$ and $\interference{\text{ES}}{}\hspace{-0.5mm}=\hspace{-0.5mm}\interference{\text{ES}}{\text{G}}+\interference{\text{IS}}{\text{A}}$, where $\interference{s}{l}$ is given by
%
%
%
%
\begin{align}  
	&
	\interference{s}{l}
	=
\sum_{
	\mathbf{x} \in \hpppI{l} \backslash  
	\left\lbrace \servingBS \right\rbrace 
	}^{}
\tpower \fading{\mathbf{x}} \ld{\hd}
G(\hd,\tilt).
\end{align}
%
%
%

Using the \ac{SINR} in \eqref{eq:SINR}, when the user associates to a $j$-group \ac{BS} with the distance $\shd$ and the tilt angle $\tilt$ under the channel environment $v$, the outage probability is given by 
\begin{align} \label{eq:def_op_SINR}
	\pout{v}{\shd}{\tilt}
=
\prob{
		\gamma^s_v(\shd,\tilt)	<	\target
	},
\end{align}
where $\target=2^{\frac{R_\text{o}}{W}}-1$ is the target \ac{SINR}.
Here, $R_\text{o}$ is the target data rate and $W$ is the bandwidth allocated to each user \cite{SalIkk:09}.
In the following theorem, we derive the network outage probability.
For readability, instead of notation $\tilt$, when scheme $s$ is used, we denote antenna tilt angles of the \ac{GBS} and the \ac{ABS} as $\tiltg$ and $\tilta$, respectively.
Note that in the \ac{IS-BS} scheme, since all \acp{BS} serve both \acp{GU} and \acp{AU}, we have a single antenna tilt angle $\tiltIS$, i.e., $\tiltgIS=\tiltaIS=\tiltIS$.
\begin{theorem} \label{thm:netout}
For \ac{IS-BS} $\left( s=\text{IS}\right)$ and \ac{ES-BS} $\left( s=\text{ES}\right)$ schemes, the network outage probability can be presented as a function of \ac{BS} antenna tilt angles $\left( \tiltg,\tilta \right) $ as
\begin{align} \label{eq:Network_Pout_general}
&\mathcal{P}_{\text{no}}^s(\tiltg,\tilta)
\!=\!
\rho_\text{G}\hspace{-3.5mm}
\underbrace{
	\sum_{\substack{ j\in \mathcal{J},\\	v\in\{\text{L},\text{N}\}	}}^{}  \hspace{-3mm}
			\left( \hspace{-1mm}
					\int_{b_{\text{G},j}(\tiltg)}^{b_{{\text{G}},{j+1}}(\tiltg)}
				\hspace{-2mm}	
				\mathcal{A}_{vj}^{a}
					\pout{v}{r}{\tiltg}
				f_{r_\text{G}^{vj}}^{s,a}(r)	dr \hspace{-1.5mm}			
			\right)
		}_{\mathcal{P}_{\text{no},\text{G}}^s(\tiltg)}	
\nonumber
\\
&\hspace{6mm}
+ \hspace{-0.5mm}
\rho_\text{A} \hspace{-3.5mm}
\underbrace{
\sum_{\substack{ j\in \mathcal{J},\\	v\in\{\text{L},\text{N}\}	}}^{} \hspace{-3mm}
\left( \hspace{-0.5mm}
		\int_{b_{\text{A},j}(\tilta)}^{b_{\text{A},j+1}(\tilta)} 
		\hspace{-1mm}
		\mathcal{A}_{vj}^{a}
		\hspace{-0.5mm}
		\pout{v}{r}{\tilta}
		f_{r_\text{A}^{vj}}^{s,a}(r)	dr	\hspace{-0.5mm}
\right) 	}_{\mathcal{P}_{\text{no},\text{A}}^s(\tilta)},
%
%
%
\end{align}
where $\mathcal{P}_{\text{no},i}^s(\tilti)$ is the network outage probability of $i$-type user for $s\in \{\text{IS},\text{ES}\}$, and $\rho_{i}=\lambda_i/\left( \lambdag+\lambdaa\right)$ is the ratio of the density of $i$-type users to that of total users, $i\in\{\text{G,A}\}$, and $f^{s,a}_{\rvr}(r)$ is given in \eqref{eq:link_pdf}.
In \eqref{eq:Network_Pout_general}, $\pout{v}{r}{\tilti}$ is given by
\begin{align}
	\label{eq:gen_outage_net} 
	&\pout{v}{r}{\tilti}
	\\
	&\hspace{-0.8mm}=	
\hspace{-0.5mm}
1
\hspace{-0.7mm}-\hspace{-1.3mm}
\sum_{n=0}^{\mfactor{v}\hspace{-0.3mm}-\hspace{-0.5mm}1}
\hspace{-1.3mm}
\left[	  \hspace{-0.8mm}
\frac{(-z)^{\hspace{-0.5mm}n}}{n!} \hspace{-0.3mm}
\frac{d^n}{dz^n} 	\hspace{-0.5mm}
\exp \hspace{-1mm}
\left( 
		\hspace{-0.5mm}	-z\noise 	\hspace{-0.3mm}
\right) 
\hspace{-1mm}
\mathcal{L}_{\interference{s}{}}(z) 	\hspace{-0.7mm}
\right]_{ z=\frac{\mfactor{v} \target}{\tpower \ld{\shd} G_j(r,\tilti)}	}.
\nonumber
\end{align}
In \eqref{eq:gen_outage_net}, $\laplace{\interference{\text{IS}}{}}=\laplace{\interference{\text{IS}}{\text{O}}}$ and $\laplace{\interference{\text{ES}}{}}=\laplace{\interference{\text{ES}}{\text{G}}}\laplace{\interference{\text{ES}}{\text{A}}}$,
where $\laplace{\interference{s}{l}}$
is the Laplace transform of the interference from $l$-type \acp{BS}, $l\in\{\text{O,G,A}\}$, for the \ac{BS} service provisioning scheme $s$, given in \eqref{eq:thm_laplace}, shown at the top of next page.
%
%
%
%
In \eqref{eq:thm_laplace}, $c_{k,j}(r,\tiltl)
=
\min\left[ 	b_{k\hspace{-0.2mm} ,j\hspace{-0.2mm} +\hspace{-0.2mm} 1 \hspace{-0.5mm}}(\tiltl),\max(r,b_{k\hspace{-0.2mm} ,j\hspace{-0.2mm}}(\tiltl))	\right]$.
%
%
%
\setcounter{eqnback}{\value{equation}}
\setcounter{equation}{24}
\begin{figure*}[t!]
	\begin{align}
		&\mathcal{L}_{I^s_l}\hspace{-0.3mm}(\hspace{-0.4mm} z\hspace{-0.4mm} ) \label{eq:thm_laplace}
		\hspace{-0.8mm}
		= \hspace{-0.7mm}
		\exp \hspace{-1.1mm}
		\left[ \hspace{-0.8mm}
		-2\pi\lambda_{I\hspace{-0.5mm},l}^s \hspace{-1.0mm}
		\sum_{j \in \mathcal{J}}^{} \hspace{-1mm}
		\left\{ \hspace{-0.5mm}
		\int_{c_{k,j} (r,\tiltl)}^{b_{k\hspace{-0.2mm} ,j\hspace{-0.2mm} +\hspace{-0.2mm} 1 \hspace{-0.5mm}}(\tiltl)} \hspace{-2.0mm} 
		t p_\text{L}(\hspace{-0.3mm}t\hspace{-0.3mm}) \hspace{-1.3mm}
		\left( \hspace{-1.0mm}
		1 \hspace{-0.5mm}-\hspace{-0.5mm}
		\frac{1}
		{\left( \hspace{-0.5mm}
			1\hspace{-0.5mm}+\hspace{-0.5mm}\frac{z}{\mfactor{\text{L}}} \tpower l_\text{L}(\hspace{-0.3mm}t\hspace{-0.3mm})  G_{j\hspace{-0.5mm}}(\hspace{-0.mm}t\hspace{-0.3mm},\hspace{-0.3mm}\tiltl \hspace{-0.3mm})
			\hspace{-0.5mm}
			\right)
			^{\hspace{-1mm} \mfactor{\text{L}}}} \hspace{-1.0mm}
		\right) \hspace{-1.2mm} dt \hspace{-1.3mm}
		\right. \right. 
		\left.\left.
		\hspace{-0.7mm}+\hspace{-1mm}
		\int_{c_{k,j} (r,\tiltl)}^{b_{k\hspace{-0.2mm} ,j\hspace{-0.2mm} +\hspace{-0.2mm} 1 \hspace{-0.5mm}}(\tiltl)}\hspace{-2.0mm}
		t p_\text{N}(\hspace{-0.3mm}t\hspace{-0.3mm}) \hspace{-1.0mm}
		\left(  \hspace{-1.0mm}
		1\hspace{-0.5mm} -\hspace{-0.5mm}
		\frac
		{1}{	1\hspace{-0.8mm}+ \hspace{-0.8mm}z \tpower l_\text{N}(\hspace{-0.3mm}t\hspace{-0.3mm})  G_{j\hspace{-0.5mm}}(\hspace{-0.3mm}t\hspace{-0.3mm},\hspace{-0.3mm}\tiltl\hspace{-0.3mm})}
		\hspace{-1mm} 
		\right) \hspace{-1mm} 
		dt \hspace{-0.5mm} 
		\right\} 
		\hspace{-0.3mm}  \right]\hspace{-0.7mm} . \hspace{-1.0mm} 
	\end{align}
	\setcounter{eqncnt}{\value{equation}}
	\centering \rule[0pt]{18cm}{0.3pt}
	\vspace{-3.0mm}
\end{figure*}
\setcounter{equation}{\value{eqnback}}
%
%
%
\end{theorem}
\setcounter{equation}{25}%

\begin{IEEEproof}
	See Appendix \ref{app:thm1}.
\end{IEEEproof}

From Theorem \ref{thm:netout}, in the general environment, we can obtain the network outage probabilities for two types of service provisioning schemes, which consider different channel fadings for \ac{LoS} and \ac{NLoS} environments.
Here, we can see that the network outage probability is affected by the main lobe service area that the \ac{BS} can serve with the strong main lobe gain, i.e., the area with distance $r^\text{{lb}}_{i}(\tilti)(=b_{k,2})$ to $r^{\text{ub}}_{i}(\tilti)(=b_{k,3})$ from a \ac{BS} (see Fig. \ref{fig:AntennaConfiguration}).
The main lobe service area is determined by the antenna tilt angle, and the effect of the antenna tilt angle on $\left| r^\text{{ub}}_{i}(\tilti)-r^\text{{lb}}_{i}(\tilti)\right| $ is presented in the following corollary.
\begin{corollary} \label{co:col1}	
For $ \angleth < \tiltg<  \frac{\pi}{2}$ and $-\frac{\pi}{2} < \tilta <-\angleth $, 
$\left| r^\text{{ub}}_{\text{G}}(\tiltg)-r^{\text{lb}}_{\text{G}}(\tiltg)\right| $ and
$\left| r^\text{{ub}}_{\text{A}}(\tilta)-r^{\text{lb}}_{\text{A}}(\tilta)\right| $ increase,
as $\tiltg$ and $\tilta$ approach $\angleth$ and $-\angleth$, respectively. 
\end{corollary}
\begin{IEEEproof}
	From \eqref{eq:lowerbound_g} and \eqref{eq:upperbound_g}, we obtain the first derivative of
	$\left| r^\text{{ub}}_{\text{G}}(\tiltg)-r^{\text{lb}}_{\text{G}}(\tiltg)\right| $ with respect to $\tiltg$ as
	\begin{align} \label{eq:diff_der1}
		\frac{\partial}{\partial\tiltg}		\hspace{-1mm}
		\left\lbrace 						\hspace{-0.5mm}
		r^\text{{ub}}_{\text{G}}\hspace{-0.5mm}(\tiltg)
		\hspace{-0.5mm}	
		-
		\hspace{-0.5mm}
		r^\text{{lb}}_{\text{G}}\hspace{-0.5mm}(\tiltg)			 \hspace{-0.5mm}
		\right\rbrace 
		\hspace{-0.5mm}
		=
		\hspace{-0.5mm} 
		\psi(\tiltg) \hspace{-1mm}
		\left(\hspace{-0.5mm}\hBS\hspace{-0.5mm}-\hspace{-0.5mm}\hGU\hspace{-0.5mm} \right)
		\hspace{-0.5mm}<\hspace{-0.5mm}0,
	\end{align}
	for $\angleth \hspace{-0.5mm}<\hspace{-0.5mm} \tiltg \hspace{-0.5mm}<\hspace{-0.5mm} \frac{\pi}{2}$,
	where $\psi(\theta)=\text{csc}^2(\theta+\angleth)-\text{csc}^2(\theta-\angleth)$.
	In \eqref{eq:diff_der1}, the inequality is obtained since $\psi(\tiltg)<0$ and $\hBS-\hGU>0$.
	From \eqref{eq:lowerbound_a} and \eqref{eq:upperbound_a}, the first derivative of $\left| r^\text{{ub}}_{\text{A}}(\tilta)-r^{\text{lb}}_{\text{A}}(\tilta)\right| $ with respect to $\tilta$ is given by
	\begin{align} \label{eq:diff_der2}
		\frac{\partial}{\partial\tilta}
		\hspace{-1mm}
		\left\lbrace  \hspace{-0.5mm}
		r^\text{{ub}}_{\text{A}}\hspace{-0.5mm}(\tilta)
		\hspace{-0.5mm}	-\hspace{-0.5mm}
		r^\text{{lb}}_{\text{A}}\hspace{-0.5mm}(\tilta) \hspace{-0.5mm}
		\right\rbrace 
		\hspace{-0.8mm}	
		&= \hspace{-0.8mm}
		\psi(\tilta) \hspace{-1mm}
		\left(\hspace{-0.5mm}\hAU\hspace{-0.5mm}-\hspace{-0.5mm}\hBS\hspace{-0.5mm} \right)	\hspace{-0.5mm}>\hspace{-0.5mm} 0,
	\end{align}
	for $\hspace{0.5mm} - \frac{\pi}{2} \hspace{-1mm}<\hspace{-0.8mm}\tilta \hspace{-0.8mm}<\hspace{-1mm}  -\angleth$.
	In \eqref{eq:diff_der2}, the inequality is obtained since $\psi(\tilta)>0$ and $\hAU-\hBS>0$. 
	Therefore, we can see that 
	$\left| r^\text{{ub}}_{\text{G}}(\tiltg)-r^{\text{lb}}_{\text{G}}(\tiltg)\right| $ and 
	$\left| r^\text{{ub}}_{\text{A}}(\tilta)-r^{\text{lb}}_{\text{A}}(\tilta)\right| $ are monotonically decreasing function and increasing function of $\tiltg$ and $\tilta$,  respectively.
\end{IEEEproof}
\begin{remark} \label{re:rem1}
From \eqref{eq:lowerbound_g}-\eqref{eq:upperbound_a} and Corollary \ref{co:col1}, 
we can see that $\left| r^\text{{ub}}_{i}(\tilti)	\hspace{-0.5mm}-\hspace{-0.5mm}	r^\text{{lb}}_{i}(\tilti)\right|$, $r^\text{{lb}}_{i}(\tilti)$, and $r^\text{{ub}}_{i}(\tilti)$  increases, as $\tiltg$ or $\tilta$ approaches $\angleth$ or $-\angleth$, respectively.
This means the main lobe service area becomes wider as $\tiltg$ or $\tilta$ approaches $\angleth$ or $-\angleth$, respectively, as also shown in Fig. \ref{fig:AntennaConfiguration}.
However, as both $r^\text{{lb}}_{i}(\tilti)$ and $r^\text{{ub}}_{i}(\tilti)$ increases, the link distance between a \ac{BS} and a user, located in the main lobe service area, becomes larger, as shown in Fig. \ref{fig:AntennaConfiguration}.
Hence, the change of the antenna tilt angle gives conflicting impacts on the network outage probability, so we need to carefully determine the antenna tilt angle to improve the network performance.
\end{remark}
\subsection{Special Case: Noise-Limited Environments} \label{subsec:Noise}
In this subsection, we consider the \ac{NTN} for \acp{UAV} where the noise power is dominant over the interference power, i.e., the noise-limited environment.

In the noise-limited environment, for given $\tilt$, the outage probability at the $k$th user is defined as
\begin{align} \label{eq:Outage_prob_SNR}
\npout{v}{\hd}{\tilt}
=
\prob{\hat{\gamma}^s_v (\hd, \tilt)	\leq	\target},
\end{align}
where $\hat{\gamma}^s_v (\hd,  \tilt)$ is obtained from \eqref{eq:SINR} by substituting $I^{s}=0$.
In the following lemma, we derive the network outage probability depending on the ratio of \acp{GU} and \acp{AU}.
\begin{lemma}\label{lemma:lem1}
In the noise-limited environment, the network outage probability can be presented by a function of \ac{BS} antenna tilt angles $\left( \tiltg,\tilta \right) $ as
\begin{align}
\label{eq:Network_Pout}
&\hat{\mathcal{P}}_{\text{no}}^s(\tiltg,\tilta)
\!=\!
\rho_\text{G}\hspace{-3.5mm}
\sum_{\substack{ j\in \mathcal{J},\\	v\in\{\text{L},\text{N}\}	}}^{}  \hspace{-3mm}
\left( \!
\int_{b_{\text{G},j}(\tiltg)}^{b_{{\text{G}},{j+1}}(\tiltg)}
\!\!	
\mathcal{A}_{vj}^{a}
\npout{v}{r}{\tiltg}
f_{r_\text{G}^{vj}}^{s,a}(r)	dr	\!	
\right)
\nonumber
\\
&
\hspace{7mm}
+ \hspace{-0.5mm}
\rho_\text{A} \hspace{-3.5mm}
\sum_{\substack{ j\in \mathcal{J},\\	v\in\{\text{L},\text{N}\}	}}^{} \hspace{-3mm}
\left( \hspace{-0.5mm}
\int_{b_{\text{A},j}(\tilta)}^{b_{\text{A},j+1}(\tilta)} 
\hspace{-1.5mm}
\mathcal{A}_{vj}^{a}
\hspace{-0.5mm}
\npout{v}{r}{\tilta}
f_{r_\text{A}^{vj}}^{s,a}(r)	dr	\hspace{-1mm}
\right) \hspace{-1mm},   \hspace{-1.5mm}
\end{align}
for $s\in \{\text{IS}, \text{ES}\}$,
where $f^{s,a}_{\rvr}(r)$  is in \eqref{eq:link_pdf}, and $\npout{v}{r}{\tilti}$ is 
given by
\begin{align} \label{eq:outage_cor}
&\npout{v}{r}{\tilti}
=
\nonumber
\\ 
&
\hspace{-1mm}	1 \hspace{-0.7mm} - \hspace{-2mm} \sum_{n=0}^{\mfactor{v} -1} \hspace{-1mm}
\frac{1}{n!}  \hspace{-1mm}
\left( \hspace{-1mm} \frac{\mfactor{v}  \target \noise }
{\tpower \ld{\shd \hspace{-0.5mm}} G_j(\hspace{-0.3mm}r\hspace{-0.3mm},\hspace{-0.3mm}\tilti\hspace{-0.3mm})} \hspace{-0.5mm}	
\right)^{\!\!\!n} 
\hspace{-1.5mm} \exp \!
\left( \hspace{-0.5mm}
\!-
\frac{\mfactor{v} \target \noise }
{\tpower \ld{\shd \hspace{-0.5mm}} G_j(\hspace{-0.3mm}r\hspace{-0.3mm},\hspace{-0.3mm}\tilti\hspace{-0.3mm})}
\!	\right) \!.
\end{align}
\end{lemma}
\begin{IEEEproof}
By substituting $\interference{s}{}\hspace{-1.2mm}= \hspace{-1.2mm}0$ and applying the \ac{CDF} of the Gamma distribution in \eqref{eq:outage_thm_L},
we obtain \eqref{eq:outage_cor}.	
By replacing $\npout{v}{\shd}{\tilti}$
in \eqref{eq:outage_integral} into $\hat{\mathcal{P}}^v_{\text{o},j}(\shd, \tilti)$  and using \eqref{eq:network_outage_SINR}, we obtain \eqref{eq:Network_Pout}.
\end{IEEEproof}
Let the optimal values of the \ac{BS} antenna tilt angle for the \ac{IS-BS} and \ac{ES-BS} schemes that minimize $\PoutIS$ and $\PoutES$ be
$\optiltIS$ and $\optilti,~i\in\{\text{G},\text{A}\}$, respectively.
For given the optimal tilt angles, in the following corollary, we compare the network outage probabilities of the \ac{IS-BS} and \ac{ES-BS} schemes, i.e.,  $\PoutISop$ and $\PoutESop$.
\begin{corollary} \label{co:col2}
When the density of \acp{BS} approaches to infinity (i.e.,~$\lambdaBS \rightarrow \infty$)
and the optimal tilt angles are used for each scheme, the network outage probability of the \ac{ES-BS} scheme is smaller than or equal to that of the \ac{IS-BS} scheme, i.e.,
\begin{align} 
\label{eq:co}
\PoutISop \geq \PoutESop.
\end{align}
\end{corollary}
\begin{IEEEproof}
When $\lambdaBS$ approaches to infinity, $\lambda^s_{\text{B,G}}$ and $\lambda^s_{\text{B,A}}$ also approach to infinity, respectively.
Hence, in \eqref{eq:link_pdf}, regardless of the service provisioning scheme, the \acp{PDF} of the horizontal distance between the $k$th user and the serving \ac{BS} become similar, i.e., $f_{\rvr}\left(r\right) \approx f^{\text{IS},a}_{\rvr}(r) \approx f^{\text{ES},a}_{\rvr}(r)$.
Substituting $f_{\rvr}\left(r\right)$ and $\npout{v}{r}{\tilti}$ into \eqref{eq:outage_integral}, and using the optimal antenna tilt angles, the network outage probabilities of $i$-type users for the \ac{IS-BS} scheme and the \ac{ES-BS} scheme, $\hat{\mathcal{P}}_{\text{no},i}^{\text{IS}} (\optiltIS )$ and $\hat{\mathcal{P}}_{\text{no},i}^{\text{ES}} (\optilti)$, can be represented as
\begin{align}
\label{eq:outage_ISES}
\hat{\mathcal{P}}_{\text{no},i}^\text{IS}\hspace{-0.3mm}(\optiltIS) 
& \hspace{-0.5mm} = \hspace{-1mm}
\sum_{\substack{ j\in \mathcal{J},\\	v\in\{\text{L},\text{N}\}	}}^{}
\hspace{-0.5mm}
\left( 
\hspace{-0.5mm}
\int_{b_{k,j}(\optiltIS)}^{b_{k,j+1}(\optiltIS)}
\hspace{-0.5mm}
\mathcal{A}_{vj}^{a}
\npout{v}{r}{\optiltIS}
f_{\rvr}(r) dr	\hspace{-0.5mm}
\right) ,
\nonumber
 \\
\hat{\mathcal{P}}_{\text{no},i}^\text{ES}\hspace{-0.3mm}(\optilti) 
& \hspace{-0.5mm} = \hspace{-1mm}
\sum_{\substack{ j\in \mathcal{J},\\	v\in\{\text{L},\text{N}\}	}}^{}
\hspace{-0.5mm}
\left( 
\hspace{-0.5mm}
\int_{b_{k,j}(\optilti)}^{b_{k,j+1}(\optilti)}
\hspace{-0.5mm}
\mathcal{A}_{vj}^{a}
\npout{v}{r}{\optilti}
f_{\rvr}(r)	dr	\hspace{-0.5mm}
\right)\hspace{-1mm}.
\end{align}
In \eqref{eq:outage_ISES}, we can always obtain $\hat{\mathcal{P}}_{\text{no},i}^{\text{IS}}\left( \optiltIS \right) \ge \hat{\mathcal{P}}_{\text{no},i}^{\text{ES}}\left( \optilti \right)$, $\forall i \in \{\text{G},\text{A}\}$
because $\optiltg$ and $\optilta$ in the \ac{ES-BS} scheme are optimized ones for \acp{GU} and \acp{AU}, respectively, while in the \ac{IS-BS} scheme, $\optiltIS$ is optimized one for both type users to minimize the total network outage probability.
Therefore, from \eqref{eq:network_outage_SINR}, we can conclude as \eqref{eq:co}. 
\end{IEEEproof}

From Corollary \ref{co:col2}, we can see that when the density of \acp{BS} is sufficiently large, the \ac{ES-BS} scheme outperforms the \ac{IS-BS} scheme in terms of the network outage probability.
Therefore, when the number of \acp{BS} is large enough, 
it is beneficial to exclusively serve \acp{GU} and \acp{AU} by independently optimizing the \ac{BS} antenna tilt angles.
This is also verified in Section \ref{sec:numerical}, through the simulation results. 

In noise-limited environments, to obtain the insight of network parameters on the network outage probability, we simplify the antenna gain model in \eqref{eq:AG} as
\begin{align}	\label{eq:AG_simple}
\tilde{G}(\hd,\tilt)=
	\begin{dcases}
		\tilde{G}_1,
		& \bound{1} < \hd < \bound{2}
		\\
		\tilde{G}_2, 
		&\bound{2}\leq \hd  \leq \bound{3}
		\\
		\tilde{G}_3,
		& \bound{3} < \hd < \bound{4}
	\end{dcases},
\end{align}
where $\tilde{G}_1=\tilde{G}_3$ is the constant antenna side lobe gain and $\tilde{G}_2$ is the constant antenna main lobe gain.
We then obtain the network outage probability as in the following corollary.
\begin{corollary} \label{co:col3}
	When $\losP{\text{N}\hspace{-0.5mm}}{\hd} \hspace{-0.5mm}=\hspace{-0.5mm}1$ and $\alpha_\text{N}=4$, the network outage probability with the simplified antenna gain model in \eqref{eq:AG_simple} is given by
\begin{align}
	\label{eq:simple_no}
	\tilde{\mathcal{P}}^s_{\text{no}} \left(\tiltg,\tilta \right) 
	=
	1- \hspace{-2mm}
	&\sum_{\substack{ j\in \mathcal{J},\\	i\in\{\text{G},\text{A}\}	}}^{}
	\hspace{-1mm}
	\rho_i
	\hspace{-1mm}
	\left\{
	g_j \hspace{-1mm}
	\left( 	b_{i,j+1}	\hspace{-0.3mm}	(\tilti)\hspace{-0.3mm}	\right) 
	\hspace{-0.5mm}-\hspace{-0.5mm}
	g_j \hspace{-1mm}
	\left( 	b_{i,j} \hspace{-0.3mm} (\tilti)\hspace{-0.3mm}	\right) 
	\hspace{-0.5mm}
	\right\}	\hspace{-1mm}	
	\nonumber \\ 
	& \times \frac
	{
		\sqrt{\hspace{-0.3mm} \tilde{G}_j \hspace{-0.3mm}} 
		\pi^{\frac{3}{2}} \hspace{-0.5mm}
		\lambda^{s}_{\text{B},i}
		\hspace{-0.5mm}
		\exp \hspace{-1.1mm}
		\left( 	\hspace{-1.3mm}
		\frac
		{
			4\omega \tilde{h}_i^2+ \pi \tilde{G}_j\lambda^{s}_{\text{B},i}   }
		{4\omega}
		\hspace{-1.3mm}	
		\right)
	}
	{2\sqrt{\omega}},
\end{align}
%
%
%
%
%
%
where $\tilde{h}_i=|\hBS-h_i|$, $\omega=\frac{\target \noise}{\tpower}$, and $g_j(b)$ is given by
	\begin{align}
		g_j(b)
		=
		\text{erf}
		\left( 
		\frac{\tilde{G}_j \lambda^{s}_{\text{B},i} \pi+ 2 \omega (\sqrt{(b^2+\tilde{h}_i^2})^2}
		{2\sqrt{\omega \tilde{G}_j}}
		\right).
	\end{align}
\end{corollary}
	\begin{IEEEproof}
From \eqref{eq:outage_cor},
by substituting  $\losP{\text{N}\hspace{-0.5mm}}{\shd} \hspace{-0.5mm}=\hspace{-0.5mm}1$,  $\alpha_\text{N}=4$, and $G(\shd,\tilti)=\tilde{G}(\shd,\tilti)$, $\tilde{\mathcal{P}}_{\text{o},j}(\shd,\tilti)$ is represented by
\begin{align} \label{eq:simple_op}
	\tilde{\mathcal{P}}_{\text{o},j}(\shd,\tilti)
	=
	1-
	\exp
		\left( -
				\frac
				{\target \noise \left(\sqrt{\shd^2+ \tilde{h}_i^2}\right) ^{4} }
				{\tpower \tilde{G}_j}
		\right) .
\end{align}
Similar to \eqref{eq:outage_integral}, after averaging $	\tilde{\mathcal{P}}_{\text{o},j}(\shd,\tilti)$ over $\shd$, we obtain the network outage probability of $i$-type user, 	$\tilde{\mathcal{P}}_{\text{no},i}(\shd,\tilti)$ as
\begin{align}
	\label{eq:simple_i_no}
			\tilde{\mathcal{P}}_{\text{no},i  \hspace{-0.8mm}} ^s
					\left(\hspace{-0.3mm}	\tilti	\hspace{-0.3mm} \right) 
			&		\hspace{-0.8mm}			= 	\hspace{-0.8mm}
			1	\hspace{-0.5mm}
			-\hspace{-1mm} 
			\sum_{\substack{ j\in \mathcal{J}}^{}} \hspace{-0.8mm}
				\int_{b_{i,j\hspace{-0.3mm}}(\tilti\hspace{-0.3mm})}^{b_{i,j+1\hspace{-0.5mm}}(\tilti\hspace{-0.3mm})}
				\hspace{-4mm}
				\exp	\hspace{-1mm}	\left( \hspace{-1mm} -
				\frac{\omega \hspace{-1mm} \left(\hspace{-1mm} \sqrt{r^2+\tilde{h}_i^2}\right) ^{\hspace{-1.3mm}4} }{\tilde{G}_j}
					\hspace{-0.5mm}	\right)  \hspace{-0.8mm}
			 \hspace{-0.5mm}
				\tilde{f}_{r_k  \hspace{-0.5mm}}^{s}(r)			 dr
			,
		\end{align}
where $\tilde{f}_{r_k}^{s}(r)\hspace{-1mm}=\hspace{-1mm}2\pi\lambda^{s}_{\text{B},i} r\exp\left(-\pi \lambda^{s}_{\text{B},i} r^2\right)$ and $\tilde{f}_{r_k}^{s}(r)\hspace{-1mm}=\hspace{-1mm}\tilde{f}_{r_i}^{s}(r)$ for $k\in\mathcal{U}_i$.
From \eqref{eq:simple_i_no}, by using result in \cite[eq. (3.322)]{tableint:07}
and \eqref{eq:network_outage_SINR}, we obtain \eqref{eq:simple_no}.
\end{IEEEproof}

In Corollary \ref{co:col3}, we obtain the network outage probability in the noise-limited environment as a closed form.

\begin{table}[!t]
	\caption{Environment parameters \label{table:parameter}} 
	\begin{center}
		\renewcommand{\arraystretch}{1.5}
		\begin{tabular}{l l | l l}
			\hline 
			{\bf Parameters} & {\bf Values} & {\bf Parameters} & {\hspace{0.32cm}}{\bf Values} \\
			\hline 
			\hspace{0.15cm}$\alpha_\text{L}$& \hspace{0.2cm}$2.5$ 
			& \hspace{0.12cm}$\alpha_\text{N}$ & \hspace{0.2cm}$3.5$ 
			\\ 	\hline 
			\hspace{0.15cm}$\tpower$ [W] & \hspace{0.2cm}$3.5$ 
			& \hspace{0.12cm}$\noise$ [W] & \hspace{0.2cm}$10^{-9}$  
			\\ 	\hline 
			\hspace{0.15cm}$\mu$& \hspace{0.2cm}$0.5$ 
			& \hspace{0.12cm}$\nu$& \hspace{0.2cm}$3\times10^{-4}$  
			\\ 	\hline 
			\hspace{0.15cm}$\xi$ & \hspace{0.2cm}$40$ 
			& \hspace{0.12cm}$\target$ & \hspace{0.2cm}$1$ 
			\\ 	\hline 
			\hspace{0.15cm}$m$ & \hspace{0.2cm}$3$ 
			& \hspace{0.12cm}$\hBS$ [m] & \hspace{0.2cm}$30$ 
			\\ 	\hline 
			\hspace{0.15cm}$\hGU$ [m] & \hspace{0.2cm}$0$ 
			& \hspace{0.12cm}$\hAU$ [m] & \hspace{0.2cm}$20$ 
			\\ 	\hline 
			\hspace{0.15cm}$\lambdaBS$ [BSs/m$^2$] & \hspace{0.2cm}$10^{-5}$ 
			& \hspace{0.12cm}$\userratio$  &\hspace{0.2cm}$0.5$ 
			\\	\hline
		\end{tabular}
	\end{center}
\end{table}%

\section{Numerical Results}\label{sec:numerical}

In this section, we evaluate the effect of the \ac{BS} antenna tilt angle, the \ac{BS} density, the interfering \ac{BS} density, and the network parameters on the network outage probability.
We first show the network outage probability on each of the \ac{IS-BS}  and \ac{ES-BS} schemes.
 We then compare the performance of the service provisioning schemes.  
In the numerical results, for the convenience of explanation, we denote the total interfering density as $\lambdaI$ regardless of the scheme, i.e., $\lambdaI
\hspace{-0.5mm}=\hspace{-0.5mm}\lambdaInt{\text{IS}}{\text{O}} \hspace{-0.5mm}=\hspace{-0.5mm}\lambdaInt{\text{ES}}{\text{G}}+\lambdaInt{\text{ES}}{\text{A}}$.
Unless otherwise specified, we use the simulation parameters given in Table \ref{table:parameter} based on the \ac{3GPP} specification and consider the dense urban environment parameters $\mu$, $\nu$ and $\xi$ \cite{3GPP:TR:36.777:V15.0.0,HolPec:08}.
\begin{figure}[t!]
	\begin{center}
		{
			\psfrag{X}[tc][bc][0.8] {Height of \ac{AU}, $\hAU$ [m]}
			\psfrag{Y}[bc][tc][0.8] {Network Outagpe Probability of \acp{AU}, $\mathcal{P}^s_{\text{no},A}(\tilta)$}
			\psfrag{AAAAAAAAAAAAAAAAAAAA}[Bl][Bl][0.65] {$h_A=\bar{h}_A$}
			\psfrag{B}[Bl][Bl][0.65] {$h_A \sim u[\bar{h}_A-5,\bar{h}_A+5]$}
			\psfrag{C}[Bl][Bl][0.65] {$h_A \sim u[\bar{h}_A-10,\bar{h}_A+10]$}	
			\psfrag{D}[Bl][Bl][0.65] {$\lambdaBS=5\times10^{-6}$ [BSs/m$^2$]}
			\psfrag{E}[Bl][Bl][0.65] {$\lambdaBS=10^{-5}$ [BSs/m$^2$]}
			\psfrag{F}[Bl][Bl][0.65] {$\lambdaBS=2\times10^{-5}$ [BSs/m$^2$]}		
		\includegraphics[width=1.0\columnwidth]{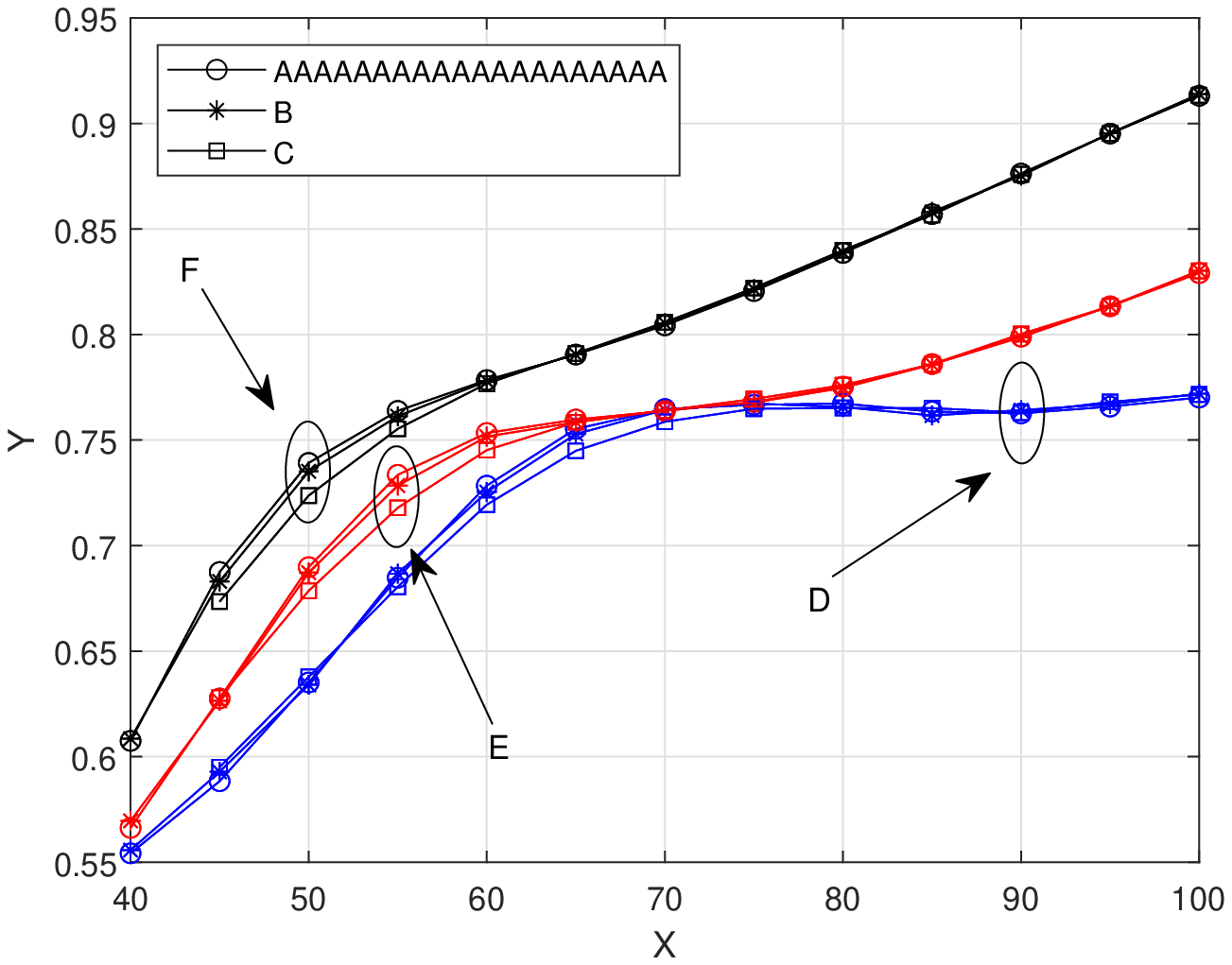}
		}
	\end{center}
	\vspace{-3.0mm}
	\caption{
		Network outage probability of the \acp{AU} as a function of $\hAU$ for different \ac{AU} height distributions and the \ac{BS} density.
		}
	\label{fig:random_height}
	\vspace{-5mm}
\end{figure}

Figure~\ref{fig:random_height} presents the network outage probability of \acp{AU} for the cases of the fixed height $\bar{h}_A$ and the uniform distribution height (i.e., $\hAU \sim u[\bar{h}_A-\delta, \bar{h}_A+\delta]$).
As shown in this figure, the trends of network outage probability with the random height are similar to that with the fixed height only.
Therefore, from this result, we show that only the performance of the fixed height case in the following figures.
Even though there is a gap between the performance of the random height and that of the fixed height, the optimal height that minimizes network outage probability is almost the same.

\subsection{Network Outage Probability of Ground and Air Users} \label{subsec:nu_general}

In this subsection, we analyze the impact of the \ac{BS} antenna tilt angle on the network outage probabilities of \acp{GU} and \acp{AU}.

\begin{figure}[t!]
	\begin{center}
		{
			\psfrag{X}[tc][bc][0.8] {BS antenna tilt angle, $\tilti$ [$^\circ$]}
			\psfrag{Y}[bc][tc][0.8] {Network Outagpe Probability, $\mathcal{P}^s_{\text{no},i}(\tilti)$}
			\psfrag{AAAAAAAAAAAAAAAAAAAAAAAAAAA}[Bl][Bl][0.65] {General Env. Nearest (Sim.)}
			\psfrag{B}[Bl][Bl][0.65] {General Env. Strongest (Sim.)}
			\psfrag{C}[Bl][Bl][0.65] {Noise-limited Env. Nearest (Sim.)}
			\psfrag{D}[Bl][Bl][0.65] {Noise-limited Env. Strongest (Sim.)}
			\psfrag{E}[Bl][Bl][0.65] {Nearest (Analy.)}
			\psfrag{F}[Bl][Bl][0.65] {$i=\text{A}$}
			\psfrag{G}[Bl][Bl][0.65] {$i=\text{G}$}
			\psfrag{H}[Bl][Bl][0.65] {(\acp{AU})}
			\psfrag{J}[Bl][Bl][0.65] {(\acp{GU})}
			\includegraphics[width=1.0\columnwidth]{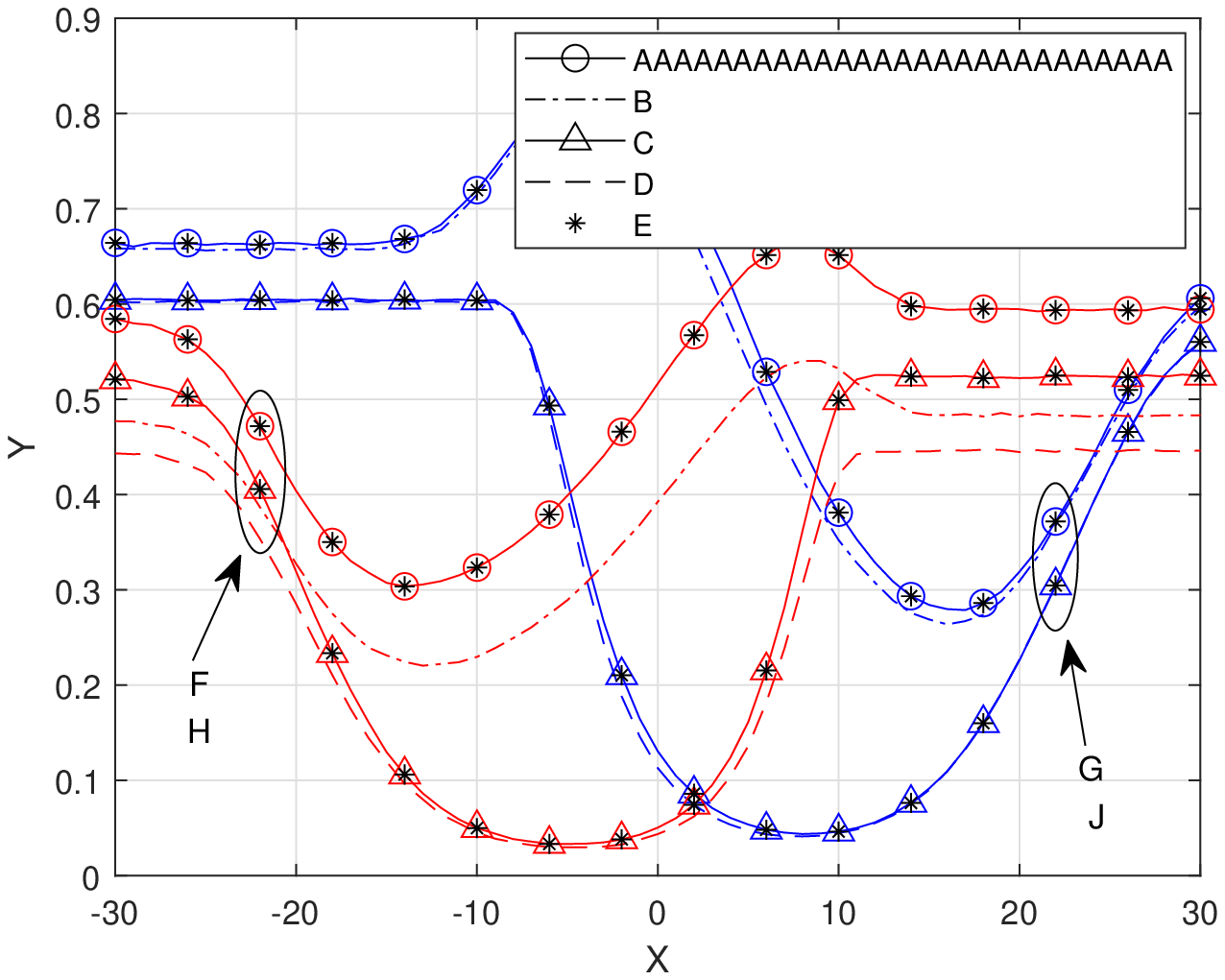}

		}
	\end{center}
	\vspace{-3.0mm}
	\caption{
		Network outage probability of $i$-type user as a function of $\tilti$ for different environments and \ac{BS} association rules. 
	}
	\label{fig:interference}
	\vspace{-5mm}
\end{figure}
Figure~\ref{fig:interference} presents the network outage probability of $i$-type user, $\mathcal{P}^s_{{\text{no},i}}(\tilti)$, as a function of the \ac{BS} antenna tilt angle, $\tilti$, for different channel environments and \ac{BS} association rules. 
Here, we use $\lambdaI=0.5\lambdaBS$.
From Fig.~\ref{fig:interference}, for \acp{GU} ($i=\text{G}$) in the general environment, we can see that as $\tilti$ increases, $\mathcal{P}^s_{{\text{no},i}}(\tilti)$ first increases up to a certain value of $\tilti$, and then decreases.
This is because as $\tilti$ increases, the number of interfering \acp{BS} that form the antenna main lobe gain to the \ac{GU} increases i.e., the \ac{GU} receives larger interference.
However, for relatively large $\tilti$ (e.g., $0^\circ<\tilti<15^\circ$), the desired \ac{BS} can transmit the signal with the antenna main lobe gain to the \ac{GU} mostly, while the number of interfering \acp{BS} with the antenna main lobe gain to the \ac{GU} decreases.
Therefore, $\mathcal{P}^s_{{\text{no},i}}(\tilti)$ decreases with $\tilti$.
Furthermore, when $\tilti$ is much large (e.g., $\tilti>20^\circ$), 
as $\tilti$ increases, the desired \ac{BS} transmits the signal with the antenna side lobe gain to the \ac{GU} with high probability.
In this case, the performance loss of the main link is dominant, so $\mathcal{P}^s_{{\text{no},i}}(\tilti)$ increases again. 
For \acp{AU} ($i=\text{A}$), the trend becomes opposite, but the reason is the same as the case of \acp{GU}. 

In the noise-limited environment, the main link channel's quality, which is affected by the antenna gain, mainly determines the network performance.
Hence, we observe that as $\tilti$ increases, $\mathcal{P}^s_{{\text{no},i}}(\tilti)$ first decreases and then increases. 
This is because as $\tilti$ increases, the main lobe of serving \ac{BS} is first closer to the user, and then get further away.

We can also see that our analysis is well matched with the simulation results.
Furthermore, the network outage probability with the strongest association rule has a similar trend to that with the nearest association rule.
The network outage probability of the nearest association is always higher than that of the strongest association.
Hence, in the following figures, we present the numerical results of the nearest association only.
%
%
%
\begin{figure}[t!]
	\begin{center}
	{
		%
		\psfrag{X}[tc][bc][0.8] {BS antenna tilt angle, $\tilti$ [$^\circ$]}
		\psfrag{Y}[bc][tc][0.8] {Network Outagpe Probability, $\mathcal{P}^s_{\text{no},i}(\tilti)$}
		\psfrag{AAAAAAAAAAAAAAAAAAAAA}[Bl][Bl][0.65] {General Env.($\lambdaI=\lambdaBS$)}
		\psfrag{B}[Bl][Bl][0.65] {General Env.($\lambdaI=0.5\lambdaBS$)}
		\psfrag{C}[Bl][Bl][0.65] {General Env.($\lambdaI=0.01\lambdaBS$)}
		\psfrag{D}[Bl][Bl][0.65] {Noise-limited Env.}
		\psfrag{E}[Bl][Bl][0.65] {$i=\text{A}$}
		\psfrag{F}[Bl][Bl][0.65] {$i=\text{G}$}
		\psfrag{G}[Bl][Bl][0.65] {(\acp{AU})}
		\psfrag{H}[Bl][Bl][0.65] {(\acp{GU})}
		\includegraphics[width=1.00\columnwidth]{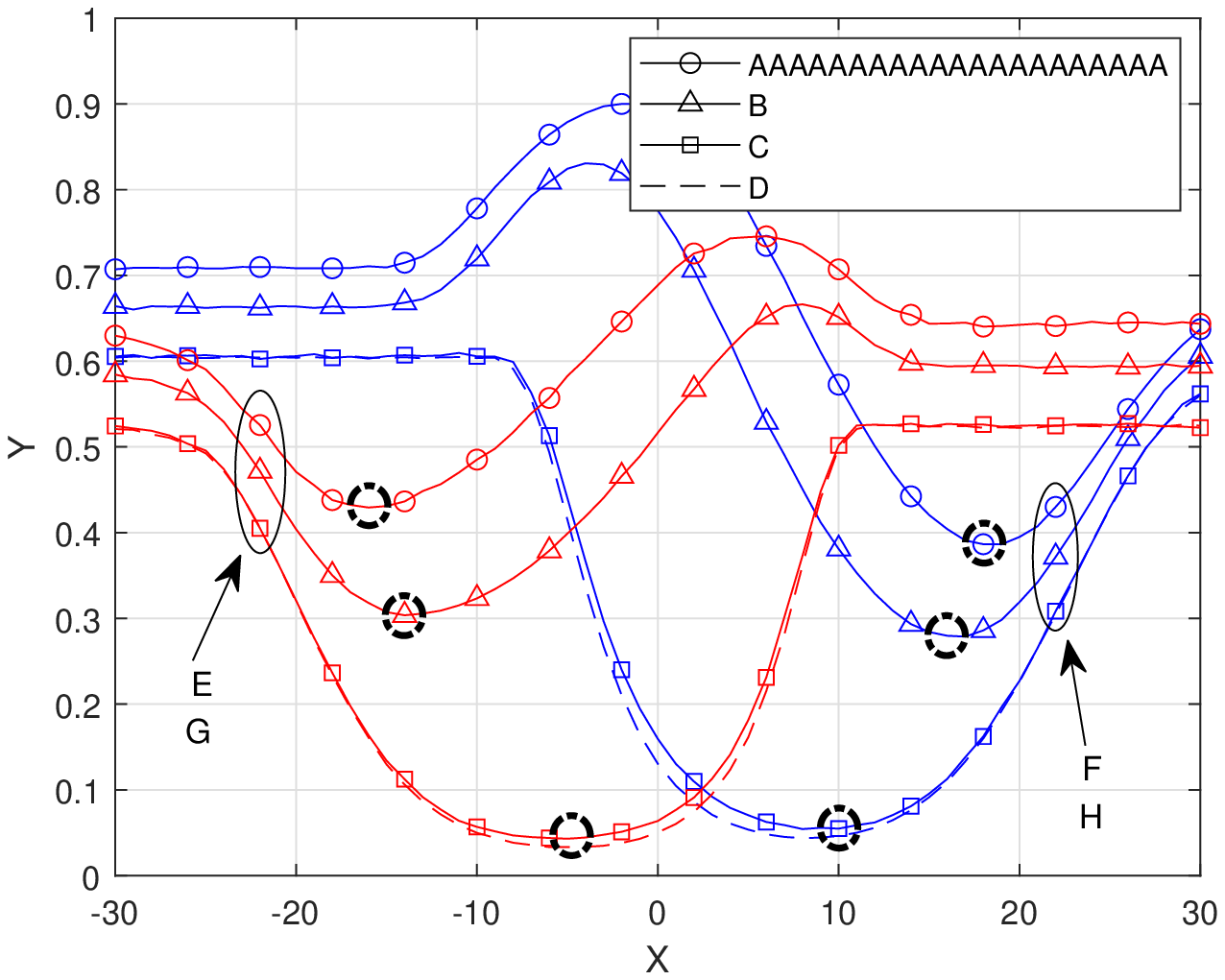}
	}
\end{center}
	\vspace{-3.0mm}
	\caption{
	Network outage probability of $i$-type user $\mathcal{P}^s_{\text{no},i}(\tilti)$ as a function of $\tilti$ for different values of $\lambdaI$. The optimal \ac{BS} antenna tilt angles,  $\optilti$, that minimize $\mathcal{P}^s_{\text{no},i}(\tilti)$ are marked by dashed circles. 
	}
	\label{fig:general_idensity}
	\vspace{-5mm}
\end{figure}

Figure~\ref{fig:general_idensity} presents the network outage probability of $i$-type user, $\mathcal{P}^s_{{\text{no},i}}(\tilti)$, as a function of the \ac{BS} antenna tilt angle, $\tilti$, for different values of the interfering \ac{BS} density, $\lambdaI$.
From Fig.~\ref{fig:general_idensity}, we can see that as $\lambdaI$ increases, the absolute value of the optimal tilt angle for $i$-type user, $\optilti$, which is marked by the dashed circle in the figure, increases.   
This is to ensure that the number of interfering \acp{BS} with the antenna main lobe gain to the \ac{GU} or \ac{AU} decreases, 
as the number of interfering \acp{BS} increases.
Moreover, when $\lambdaI$ is much small (e.g., $\lambdaI \leq 0.01\lambdaBS$), 
we can also observe that the network outage probability in the general environment approaches that in the noise-limited environment.

%
%
%
\subsection{Results of \ac{IS-BS} Scheme}

In this subsection, 
we analyze the impact of the \ac{BS} antenna tilt angle on the network outage probability with the \ac{IS-BS} scheme.


\begin{figure}[t!]
	\begin{center}
		{
			%
			\psfrag{X}[tc][bc][0.8] {BS antenna tilt angle, $\tiltIS$ [$^\circ$]}
			\psfrag{Y}[bc][tc][0.8] {Network Outagpe Probability, $\mathcal{P}^\text{IS}_{\text{no}}(\tiltIS)$}
			\psfrag{AAAAAAAAAAAAAAAAAAAAA}[Bl][Bl][0.65] {$\hBS=20$ [m], $\hAU=50$ [m]}
			\psfrag{B}[Bl][Bl][0.65] {$\hBS=30$ [m], $\hAU=40$ [m]}
			\psfrag{C}[Bl][Bl][0.65] {$\hBS=30$ [m], $\hAU=50$ [m]}
			\psfrag{D}[Bl][Bl][0.65] {$\hBS=30$ [m], $\hAU=60$ [m]}
			\psfrag{E}[Bl][Bl][0.65] {$\hBS=40$ [m], $\hAU=50$ [m]}
			\includegraphics[width=1.00\columnwidth]{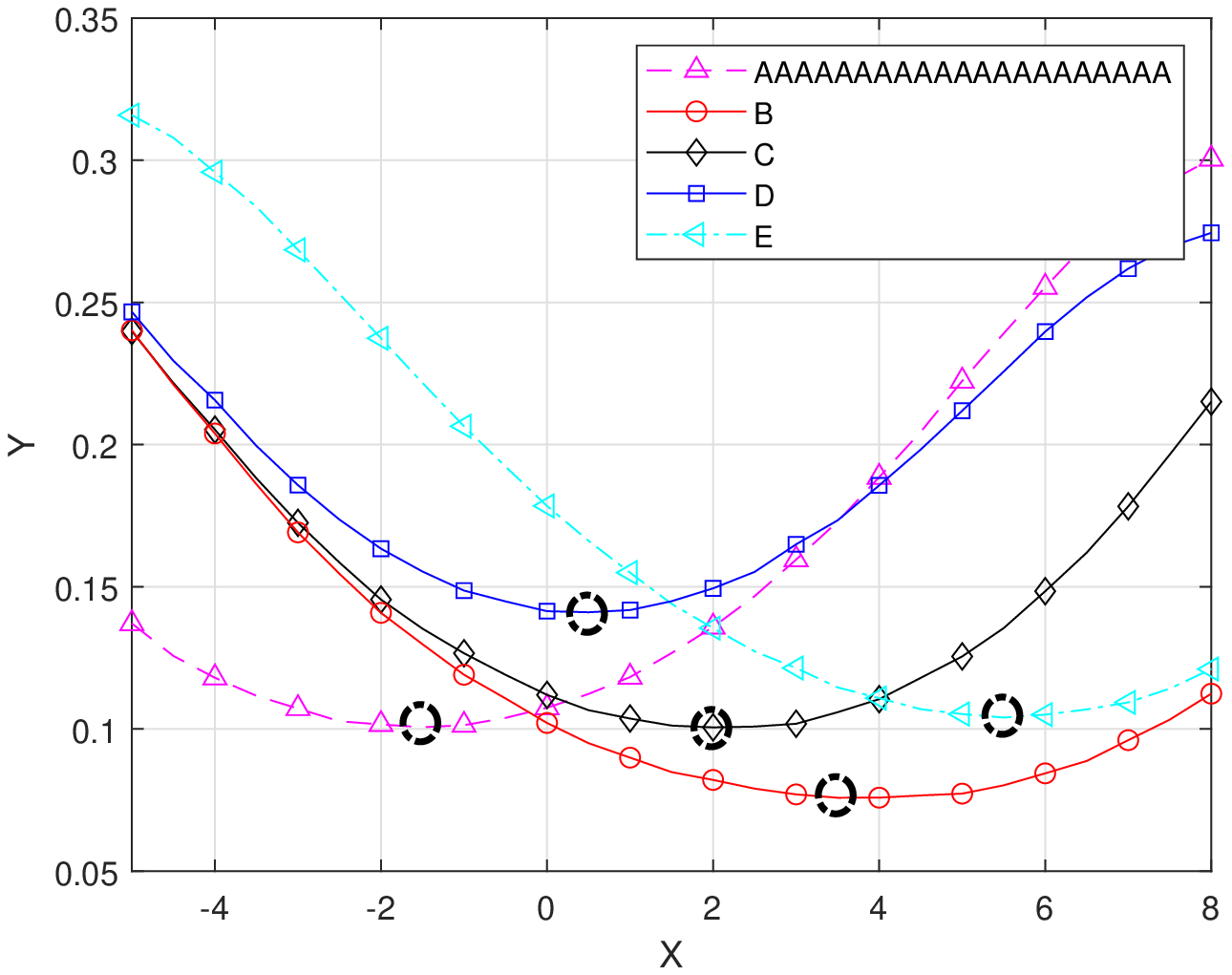}
		}
	\end{center}
	\vspace{-3.0mm}
	\caption{
		Network outage probability $\mathcal{P}^\text{IS}_{\text{no}}(\tiltIS)$ as a function of $\tiltIS$ for different values of $\hBS$ and $\hAU$ with $\lambdaI=0.01\lambdaBS$.
		The optimal \ac{BS} antenna tilt angles, $\optiltIS$, that minimize $\mathcal{P}^\text{IS}_{\text{no}}(\tiltIS)$ are marked by dashed circles.  
	}
	\label{fig:IS_tilt_pout_height}
\end{figure}

%
%
%


Figure~\ref{fig:IS_tilt_pout_height} presents the network outage probability of the \ac{IS-BS} scheme, $\mathcal{P}^\text{IS}_\text{no}(\tiltIS)$, as a function of the \ac{BS} antenna tilt angle, $\tiltIS$, for different values of the \ac{BS} height, $\hBS$, and the \ac{AU} height, $\hAU$.
Here, we use $\lambdaI=0.01\lambdaBS$ (i.e., similar to the noise-limited environment).
From Fig.~\ref{fig:IS_tilt_pout_height}, we can see that for the fixed height of \acp{AU} (e.g., $\hAU=50$~m), as the height of the \ac{BS} increases (e.g., $\hBS=20\sim40$~m),
the optimal value of the \ac{BS} antenna tilt angle, $\optiltIS$, which is marked by the dashed circle in the figure, increases.
For \acp{AU}, as $\hBS$ increases, the \ac{LoS} probability between the \ac{BS} and the \ac{AU} increases and the distance-dependent path loss decreases. 
Hence, the performance of \acp{AU} can be significantly improved by the high \ac{LoS} probability and low path loss.
On the other hand, for \acp{GU}, as $\hBS$ increases, 
the \ac{LoS} probability between the \ac{BS} and the \ac{GU} increases,
while the distance-dependent path loss increases due to the increased distance from the \ac{BS} to the \ac{GU} and it is harmful to the \ac{GU}.
Consequently, as $\hBS$ increases, since \acp{GU} experience relatively worse channel condition compared to \acp{AU}, the optimal values of the \ac{BS} antenna tilt angle increases to downward to compensate the performance loss of \acp{GU}.

In this figure, we can also observe that for the fixed height of \acp{BS} (e.g. $\hBS=30$ m), as the height of the \ac{AU} increases (e.g. $\hAU=40\sim60$ m),
the minimum network outage probability, which is a value of the dashed circle in the $y$-axis, increases and the optimal value of the \ac{BS} antenna tilt angle, which is a value of the dashed circle in the $x$-axis, decreases.
As $\hAU$ increases, the \ac{LoS} probability between the \ac{BS} and \ac{AU} and the distance-dependent path loss increases. 
However, since the effect of the path-loss increasing is greater, the outage probability of \ac{AU} increases. 
On the other hand, the performance of \acp{GU} is not affected by $\hAU$.
Therefore, the optimal antenna tilt angle decreases to be compensated for the performance loss of \acp{AU}.


\begin{figure}[t!]
	\begin{center}
		{
			%
			\psfrag{X}[tc][bc][0.8] {BS antenna tilt angle, $\tiltIS$ [$^\circ$]}
			\psfrag{Y}[bc][tc][0.8] {Network Outagpe Probability, $\mathcal{P}^\text{IS}_{\text{no}}(\tiltIS)$}
			\psfrag{AAAAAAAAAAAAAA}[Bl][Bl][0.65] {$\hBS=20$, $\hAU=50$}
			\psfrag{B}[Bl][Bl][0.65] {$\hBS=30$, $\hAU=40$}
			\psfrag{C}[Bl][Bl][0.65] {$\hBS=30$, $\hAU=50$}
			\psfrag{D}[Bl][Bl][0.65] {$\hBS=30$, $\hAU=60$}
			\psfrag{E}[Bl][Bl][0.65] {$\hBS=40$, $\hAU=50$}
			\includegraphics[width=1.00\columnwidth]{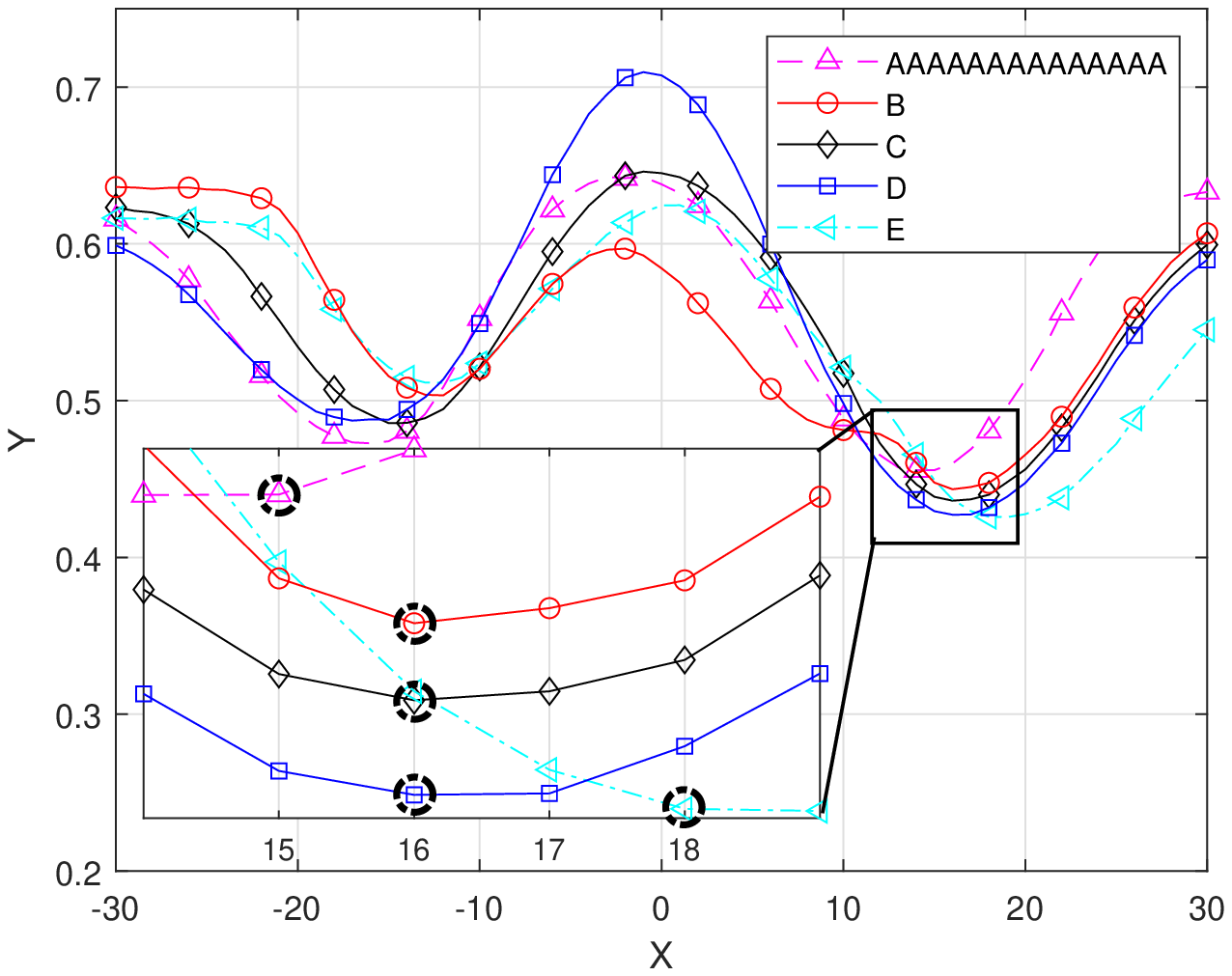}
		}
	\end{center}
	\vspace{-3.0mm}
	\caption{
		Network outage probability $\mathcal{P}^\text{IS}_{\text{no}}(\tiltIS)$ as a function of $\tiltIS$ for different values of $\hBS$ and $\hAU$ with $\lambdaI=0.5\lambdaBS$.
		The optimal \ac{BS} antenna tilt angles, $\optiltIS$, that minimize $\mathcal{P}^\text{IS}_{\text{no}}(\tiltIS)$ are marked by dashed circles.  
	}
	\label{fig:IS_tilt_pout_height_general}
\end{figure}

Figure~\ref{fig:IS_tilt_pout_height_general} presents the network outage probability of the \ac{IS-BS} scheme, $\mathcal{P}^\text{IS}_\text{no}(\tiltIS)$, as a function of the \ac{BS} antenna tilt angle, $\tiltIS$, for different values of the \ac{BS} height, $\hBS$, and the \ac{AU} height, $\hAU$, similar to Fig.~\ref{fig:IS_tilt_pout_height}.
Here, we use $\lambdaI=0.5\lambdaBS$.
From Fig.~\ref{fig:IS_tilt_pout_height_general}, we can see that the optimal values of the \ac{BS} antenna tilt angle, $\optiltIS$, exist in the considerably down tilted regions compared to Fig.~\ref{fig:IS_tilt_pout_height}.
As shown in Fig.~\ref{fig:general_idensity}, the difference of the optimal antenna tilt angles for \acp{GU} and \acp{AU} increases as $\lambdaI$ increases.
Consequently, in terms of the network performance of the \ac{IS-BS} scheme, it is worth optimizing the antenna tilt angle toward a certain type of users, i.e., \acp{GU} and \acp{AU}.
Specifically, for a given configuration, \acp{AU} are more affected by interference due to high \ac{LoS} probability than \acp{GU}, hence \acp{BS} transmit the signal to \acp{AU} with the side lobe to reduce interfering signal power.
On the other hand, to increase the main link power, \acp{BS} transmit the signal to \acp{GU} with the main lobe.
Therefore, to minimize network outage probability, the \ac{BS} antenna needs to be tilted downwards.

We can also see that for the fixed height of \acp{AU} (e.g., $\hAU=50$~m), as the height of the \ac{BS} increases (e.g., $\hBS\!=\!20\!\sim\!40$ m), the optimal value of the \ac{BS} antenna tilt angle increases.
This is to reduce the number of interfering \acp{BS} which has the antenna main lobe gain to \acp{GU} and ensure that most serving \ac{BS} transmits the signal with the antenna main lobe gain to \acp{GU}.  
On the contrary, for the fixed height of \acp{BS} (e.g., $\hBS=30$ m), there is no change in the value of optimal tilt angle according to $\hAU$, because \acp{AU} are served by the side lobe.

\begin{figure}[t!]
	\begin{center}
		{
			%
			\psfrag{X}[tc][bc][0.8] {Ratio of \acp{GU} to total users, $\userratio$}
			\psfrag{Y}[bc][tc][0.8] {Optimal \ac{BS} antenna tilt angle, $\optiltIS$ [$^\circ$]}
			\psfrag{AAAAAAAAAAAAAAAAAAA}[Bl][Bl][0.65] {$\lambdaBS=5\times10^{-6}$ [BSs/m$^2$]}
			\psfrag{B}[Bl][Bl][0.65] {$\lambdaBS=10^{-5}$ [BSs/m$^2$]}
			\psfrag{C}[Bl][Bl][0.65] {$\lambdaBS=2\times10^{-5}$ [BSs/m$^2$]}
	\includegraphics[width=1.00\columnwidth]{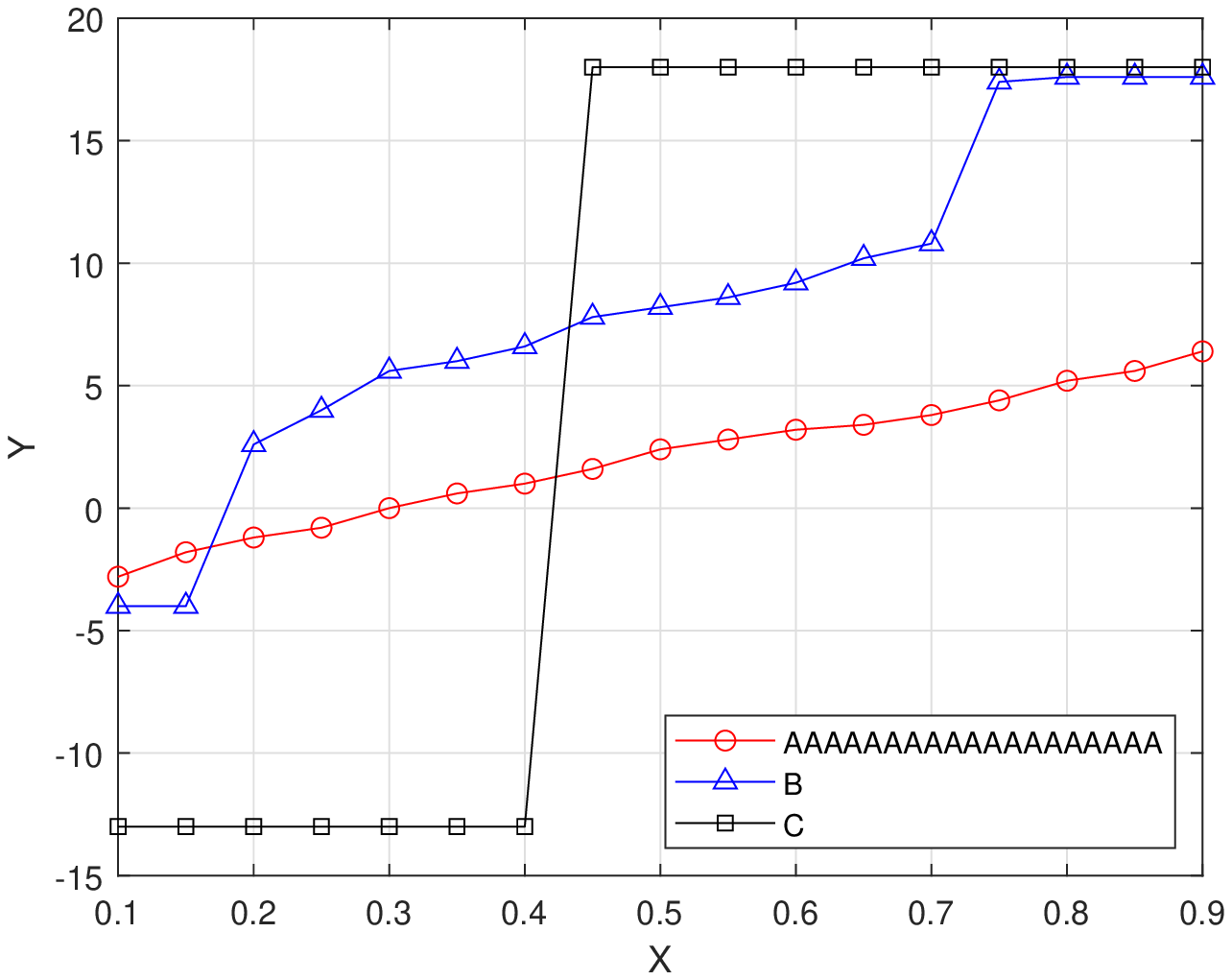}
		}
	\end{center}
	\vspace{-3.0mm}
	\caption{
		Optimal \ac{BS} antenna tilt angle, $\optiltIS$, according to $\userratio$ for different values of $\lambdaBS$ in the \ac{IS-BS} scheme.
	}
	\label{fig:IS_userdensity_opttilt}
\end{figure}

Figure~\ref{fig:IS_userdensity_opttilt} presents the optimal value of the \ac{BS} antenna tilt angle, $\optiltIS$, according to the ratio of \acp{GU} to total users, $\userratio$, for different values of the total \ac{BS} density, $\lambdaBS$,  with the \ac{IS-BS} scheme.
Here, we use $\lambdaI=0.1\lambdaBS$.
From Fig.~\ref{fig:IS_userdensity_opttilt}, we can see that as $\userratio$ increases, the optimal value of the \ac{BS} antenna tilt angle, $\optiltIS$, also increases.
Since the interference is not significant in this environment, the main link channel's quality mainly determines the network performance dominantly.
Hence, as the portion of \acp{GU} increases, the \ac{BS} needs to tilt its antenna downward. 
For large $\lambdaBS$ (e.g., $\lambdaBS \geq 2\times10^{-5}$), we can also observe that the value of the optimal antenna tilt angle is either downwards (e.g., $\optiltIS=18^\circ$) or upwards (e.g., $\optiltIS=-13^\circ$).
Because of the significant difference of the optimal antenna tilt angles for \acp{GU} and \acp{AU}, it is worth optimizing the antenna tilt angle toward a certain type of users, as also explained in Fig.~\ref{fig:IS_tilt_pout_height_general}.

\subsection{Results of \ac{ES-BS} Scheme}

In this subsection, we analyze the impact of the \ac{BS} antenna tilt angle on the network outage probability with the \ac{ES-BS} scheme.
Note that, in the \ac{ES-BS} scheme, since the \acp{GU} and \acp{AU} are exclusively served by the \acp{BS}, the antenna tilt angles for \acp{GU}, $\tiltgES$, and \acp{AU}, $\tiltaES$, are independently designed to minimize the network outage probability.
Furthermore, in the \ac{ES-BS} scheme, the ratio of \acp{GBS} affects the optimal \ac{BS} tilt angles
and hence, we optimize the \ac{BS} tilt angles in accordance with the ratio of \acp{GBS} to total \acp{BS}, $\bsratio$.

\begin{figure}[t!]
	\begin{center}
		{
			\psfrag{X}[tc][bc][0.8] {Ratio of \acp{GU} to total users$, \userratio$}
			\psfrag{Y}[bc][tc][0.8] {Optimal ratio of \acp{BS} for \ac{GU}, $\bsratio^{*}$}
			\psfrag{AAAAAAAAAAAAAAAAAAA}[Bl][Bl][0.65] {$\lambdaBS=5\times10^{-6}$ [BSs/m$^2$]}
			\psfrag{B}[Bl][Bl][0.65] {$\lambdaBS=10^{-5}$ [BSs/m$^2$]}
			\psfrag{C}[Bl][Bl][0.65] {$\lambdaBS=2\times10^{-5}$ [BSs/m$^2$]}
			
\includegraphics[width=1.00\columnwidth]{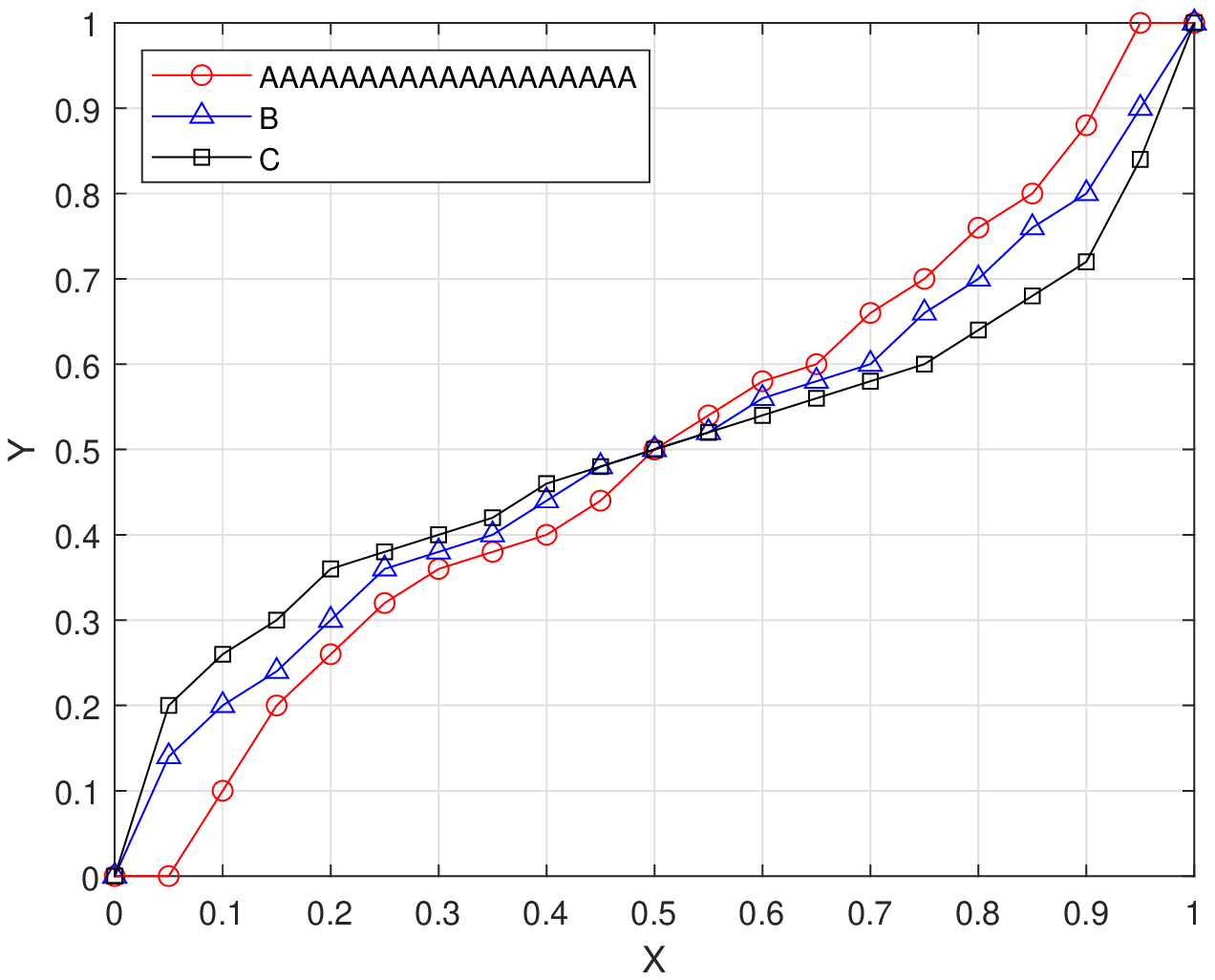}
		}
	\end{center}
	\vspace{-3.0mm}
	\caption{
		Optimal ratio of \acp{BS} for \acp{GU} to total \ac{BS} $\bsratio$ according to $\userratio$ for different values of $\lambdaBS$ in the \ac{ES-BS} scheme.
		}
	\label{fig:ES_userratio_optbsratio}
\end{figure}

Figure~\ref{fig:ES_userratio_optbsratio} shows the optimal ratio of \acp{GBS} to total \acp{BS}, $\bsratio^*$, that minimizes the network outage probability, according to the \ac{GU} ratio to total users, $\userratio$. 
We consider different values of the total \ac{BS} density, $\lambdaBS$, and we use $\lambdaI=0.1\lambdaBS$.
Here, for given $\userratio$, $\lambdaBS$, and $\bsratio$, the \ac{BS} antenna tilt angles $\tiltgES$ and $\tiltaES$ are also optimized to minimize the network outage probability.
In Fig.~\ref{fig:ES_userratio_optbsratio}, as $\userratio$ increases, $\bsratio^*$ also increases because it is beneficial to have more \acp{GBS} when the portion of \acp{GU} is large.
We can also see that for large $\userratio$ (e.g., $\userratio > 0.5$), $\bsratio^*$ becomes smaller as $\lambdaBS$ increases.
This is because as there is more the number of \acp{BS}, we can have a sufficient number of 
\acp{GBS}, so we can assign a larger portion of \acp{BS} as the \acp{ABS}.
On the contrary, when $\userratio$ is small (e.g., $\userratio < 0.5$ ), $\bsratio^*$ becomes larger as $\lambdaBS$ increases for a similar reason.

\begin{figure}[t!]
\begin{center}
{
	%
	\psfrag{X}[tc][bc][0.8] {Ratio of \acp{GU} to total users, $\userratio$}
	\psfrag{Y}[bc][tc][0.8] {Optimal \ac{BS} antenna tilt angle, $\optilti$ [$^\circ$]}
	\psfrag{AAAAAAAAAAAAAAAAAAA}[Bl][Bl][0.65] {$\lambdaBS=5\times10^{-6}$ [BSs/m$^2$]}
	\psfrag{B}[Bl][Bl][0.65] {$\lambdaBS=10^{-5}$ [BSs/m$^2$]}
	\psfrag{C}[Bl][Bl][0.65] {$\lambdaBS=2\times10^{-5}$ [BSs/m$^2$]}
	\psfrag{D}[Bl][Bl][0.65] {$\optiltg$}
	\psfrag{E}[Bl][Bl][0.65] {$\optilta$}
	\includegraphics[width=1.00\columnwidth]{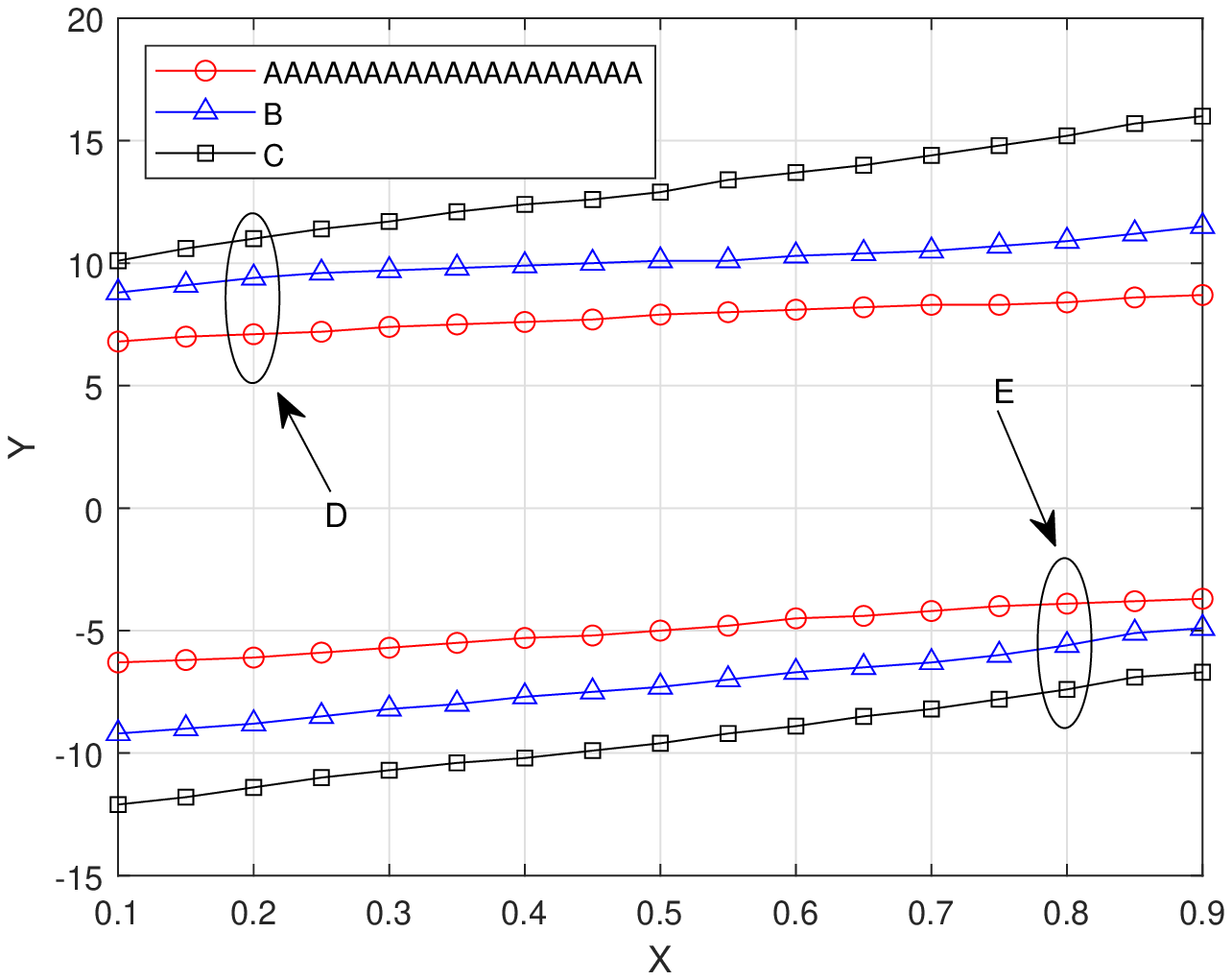}
}
\end{center}
\vspace{-3.0mm}
\caption{
		Optimal \ac{BS} antenna tilt angle ($\optiltg$, and $\optilta$) according to $\userratio$ for different values of $\lambdaBS$ in the \ac{ES-BS} scheme.
		}
\label{fig:ES_userdensity_opttilt}
\end{figure}

Figure~\ref{fig:ES_userdensity_opttilt} presents the optimal value of the \ac{BS} antenna tilt angles, $\optiltg$ and $\optilta$, according to the ratio of \acp{GU} to total users, $\userratio$, for different values of the total \ac{BS} density, $\lambdaBS$, with \ac{ES-BS} scheme.
Here, we use $\lambdaI=0.1\lambdaBS$.
From Fig.~\ref{fig:ES_userdensity_opttilt}, 
we can see that as $\userratio$ increases, the absolute values of $\optiltg$ and $\optilta$ also increase.
In the \ac{ES-BS} scheme, as $\userratio$ increases $\bsratio^*$ also increases, as shown in Fig.~\ref{fig:ES_userratio_optbsratio}. 
Therefore, as the number of \acp{BS} increases, to reduce the number of interfering \acp{BS} giving the large interference with the antenna main lobe gain, the antenna is tilted more downwards or upwards. 
For the same reason, we can observe that for given $\userratio$, as $\lambdaBS$ increases, the absolute values of $\optiltg$ and $\optilta$ also increase.



\subsection{Comparison between \ac{IS-BS} Scheme and \ac{ES-BS} Scheme}

In this subsection,  we compare the performance of the \ac{BS} service provisioning schemes in terms of the network outage probability according to the ratio of \acp{GU} to total users, $\userratio$.
As a baseline scheme, we also plot the service provisioning scheme that the \ac{BS} antenna is tilted toward \acp{GU} without considering \acp{AU} as in conventional cellular networks.
In this baseline scheme, the \ac{BS} optimizes the antenna tilt angle to minimize the outage probability of \acp{GU}.
For the comparison of the \ac{IS-BS} scheme and the \ac{ES-BS} scheme, the antenna tilt angle of the \ac{IS-BS} scheme ($\tiltIS$), that of the \ac{ES-BS} scheme ($\tiltgES,\tiltaES$), and the ratio of \acp{GBS} ($\bsratio$) are optimized, respectively.

\begin{figure}[t!]
	\begin{center}
		{
			\psfrag{X}[tc][bc][0.8] {Ratio of \acp{GU} to total users, $\userratio$}
			\psfrag{Y}[bc][tc][0.8] {Network Outagpe Probability, $\mathcal{P}^s_\text{no}(\tiltg,\tilta)$}
			\psfrag{AAAAAAAAAAAAAA}[Bl][Bl][0.65] {$\lambdaI=0.1\lambdaBS$}
			\psfrag{B}[Bl][Bl][0.65] {$\lambdaI=0.05\lambdaBS$}
			\psfrag{C}[Bl][Bl][0.65] {$\lambdaI=0.01\lambdaBS$}\begin{flushleft}
				
			\end{flushleft}
			\psfrag{D}[Bl][Bl][0.65] {Noise-limited Env.}
			\psfrag{E}[Bl][Bl][0.65] {ES-BS scheme}
			\psfrag{F}[Bl][Bl][0.65] {IS-BS scheme}
				\includegraphics[width=1.00\columnwidth]{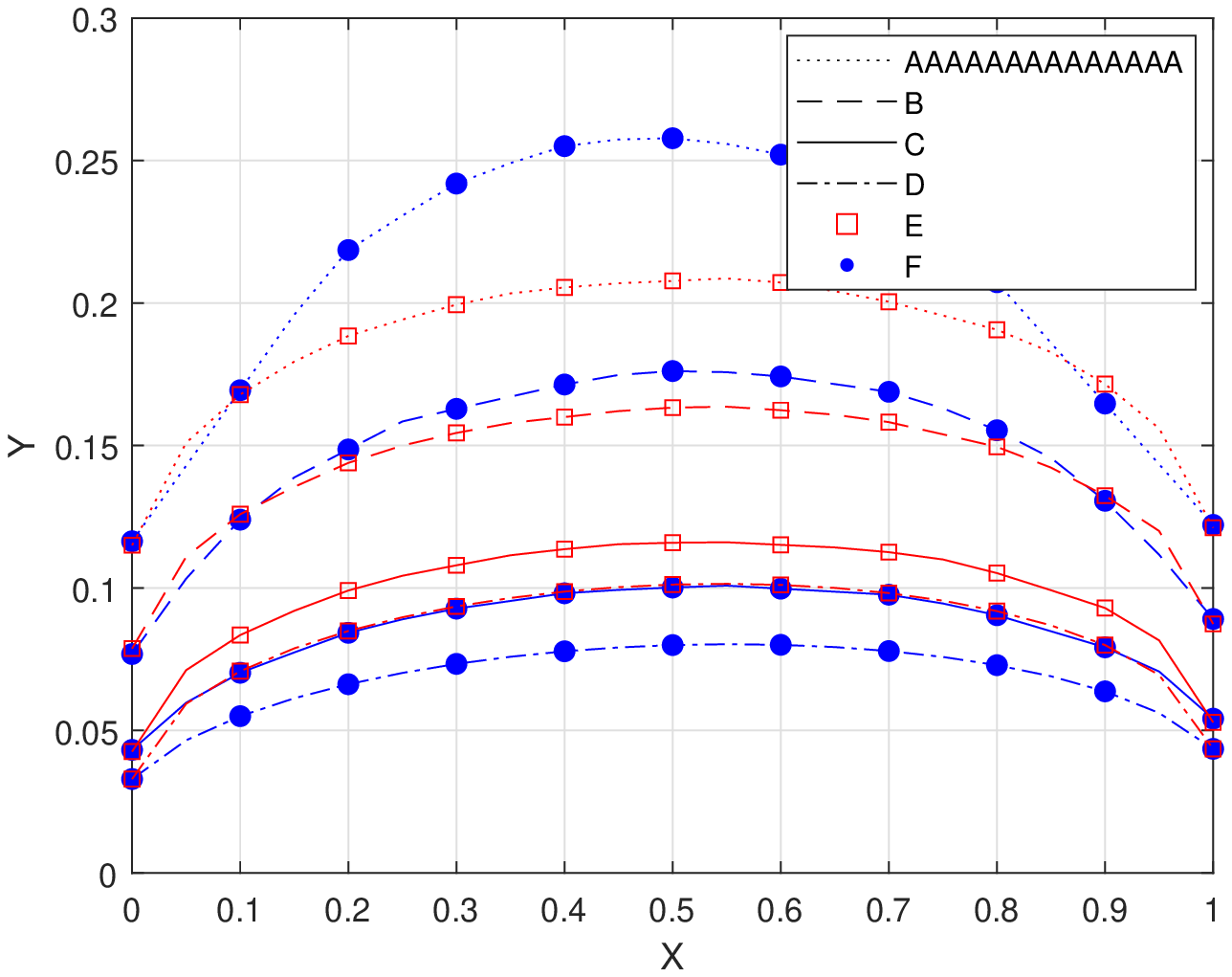}
		}
	\end{center}
	\vspace{-4mm}
	\caption{
		Network outage probability according to $\userratio$ with different values of $\lambdaI$ of \ac{IS-BS} scheme and \ac{ES-BS} scheme.
	}
	\label{fig:ESIS_general}
	\vspace{-3mm}
\end{figure}

Figure~\ref{fig:ESIS_general} presents the network outage probability, $\mathcal{P}^s_\text{no}(\tiltg,\tilta)$, as a function of the ratio of \acp{GU}, $\userratio$, for different values of $\lambdaI$ and service provisioning schemes.
From Fig.~\ref{fig:ESIS_general}, we can see that when $\lambdaI$ is large (e.g., $\lambdaI\geq0.05\lambdaBS$), the \ac{ES-BS} scheme  outperforms the \ac{IS-BS} scheme.
This is because, for the \ac{ES-BS} scheme, most of the interference from other types of \acp{BS} mostly transmits the signal to user with antenna side lobe gain.
On the other hand, for small $\lambdaI$ (e.g., $\lambdaI \leq 0.01\lambdaBS$) and the noise-limited environment, the \ac{IS-BS} scheme performs better than the \ac{ES-BS} scheme.
This is because the effect of the interference is relatively small, so more serving \acp{BS} candidates (i.e., $\lambda^\text{IS}_{\text{B},i} >\lambda^\text{ES}_{\text{B},i}$) improve the performance of the main link.

\begin{figure}[t!]
\begin{center}
{
	\psfrag{X}[tc][bc][0.8] {Ratio of \acp{GU} to total users, $\userratio$}
	\psfrag{Y}[bc][tc][0.8] {Network Outage probability, $\PoutS$}
	\psfrag{AAAAAAAA}[Bl][Bl][0.65] {ES-BS}
	\psfrag{B}[Bl][Bl][0.65] {IS-BS}
	\psfrag{C}[Bl][Bl][0.65] {Baseline}
	\psfrag{D}[Bl][Bl][0.65] {$\lambdaBS=5\times10^{-6}$ [BSs/m$^2$]}
	\psfrag{E}[Bl][Bl][0.65] {$\lambdaBS=10^{-5}$ [BSs/m$^2$]}
	\psfrag{F}[Bl][Bl][0.65] {$\lambdaBS=2\times10^{-5}$ [BSs/m$^2$]}
	\includegraphics[width=1.00\columnwidth]{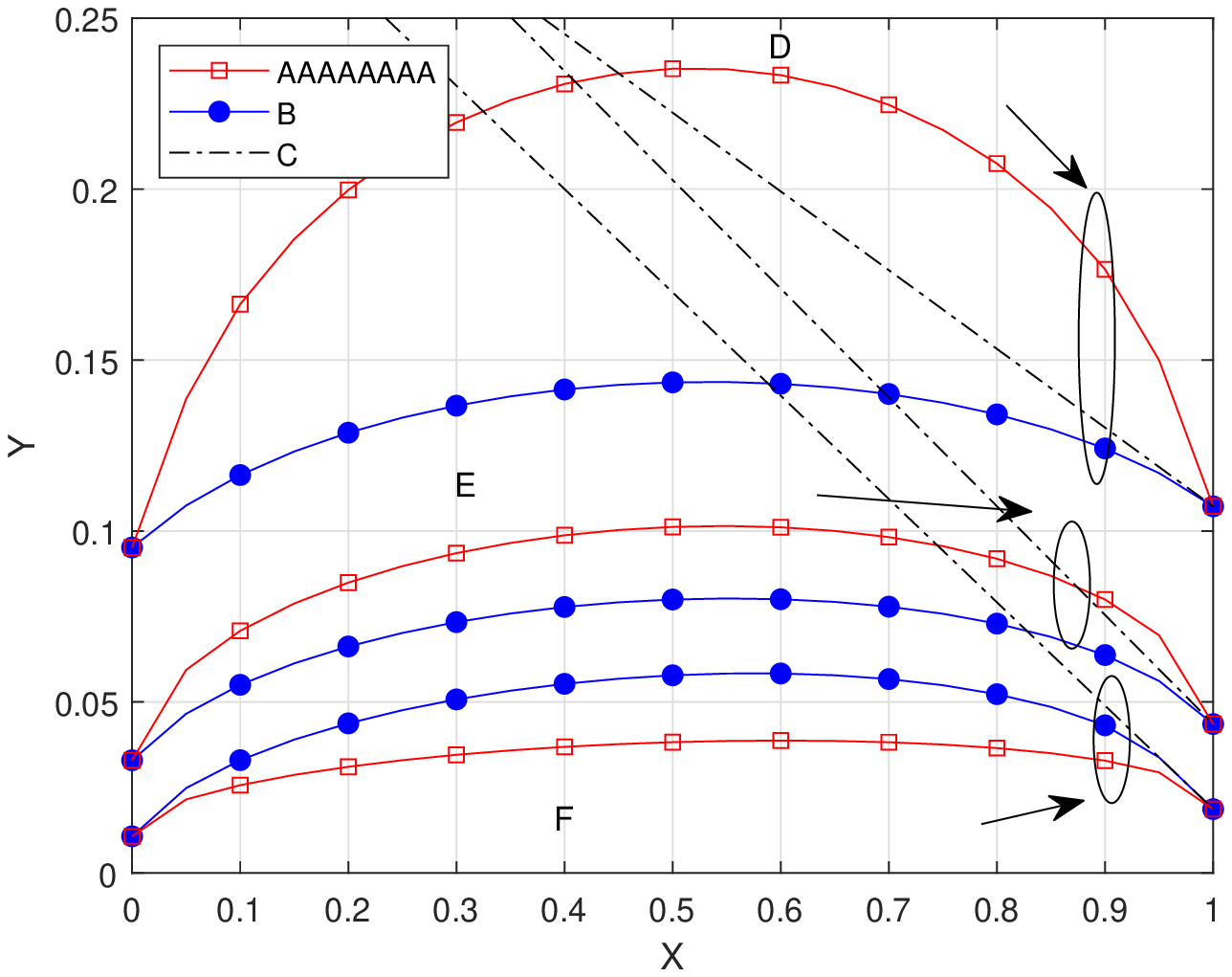}
}
\end{center}
\vspace{-3.0mm}
\caption{
	Network outage probability according to the \ac{GU} ratio with different total \ac{BS} density $\lambdaBS$ of \ac{IS-BS} scheme, \ac{ES-BS} scheme, and baseline scheme
	in the noise-limited environment.
}
\label{fig:ESIS_userratio_pout}
\end{figure}

Figure \ref{fig:ESIS_userratio_pout} presents the network outage probability in noise-limited environments, $\hat{\mathcal{P}}^s_\text{no}(\tiltg,\tilta)$, as a function of the ratio of \acp{GU} $\userratio$ for different values of the total \ac{BS} density $\lambdaBS$ and different service provisioning schemes.
From Fig.~\ref{fig:ESIS_userratio_pout}, we can see that when the total \ac{BS} density is small (e.g., $\lambdaBS \leq 10^{-5}$),
the \ac{IS-BS} scheme outperforms the \ac{ES-BS} scheme.
On the contrary, for the large total \ac{BS} density (e.g., $\lambdaBS \geq 2\times10^{-5}$), the \ac{ES-BS} scheme provides better performance than the \ac{IS-BS} scheme in terms of the network outage probability.
From these observations, we can find that when there exist enough \acp{BS} in the network, it is beneficial to exclusively serve each type of user by independently optimizing the \ac{BS} antenna tilt angle for each type of user (\ac{ES-BS} scheme).
On the other hand, when the number of \acp{BS} is relatively small, the efficient service provisioning scheme is that all \acp{BS} serve both \acp{GU} and \acp{AU} by optimizing the \ac{BS} antenna tilt angle to maximize the network performance (\ac{IS-BS} scheme).
We can also see that regardless of $\lambdaBS$ and $\userratio$, the service provisioning schemes outperform the baseline scheme because the schemes design the \ac{BS} antenna tilt angle by considering \acp{AU} as well as \acp{GU}.

\begin{figure}[t!]
	\begin{center}
		{
			\psfrag{X}[tc][bc][0.8] {\ac{BS} height, $\hBS$}
			\psfrag{Y}[bc][tc][0.8] {Critical \ac{BS} density, $\lambdaBS^\text{c}$}
			\psfrag{AAAAAAAAA}[Bl][Bl][0.65] {$\hAU=50 [m]$}
			\psfrag{B}[Bl][Bl][0.65] {$\hAU=60 [m]$}
			\psfrag{C}[Bl][Bl][0.65] {$\hAU=70 [m]$}
			\includegraphics[width=1.00\columnwidth]{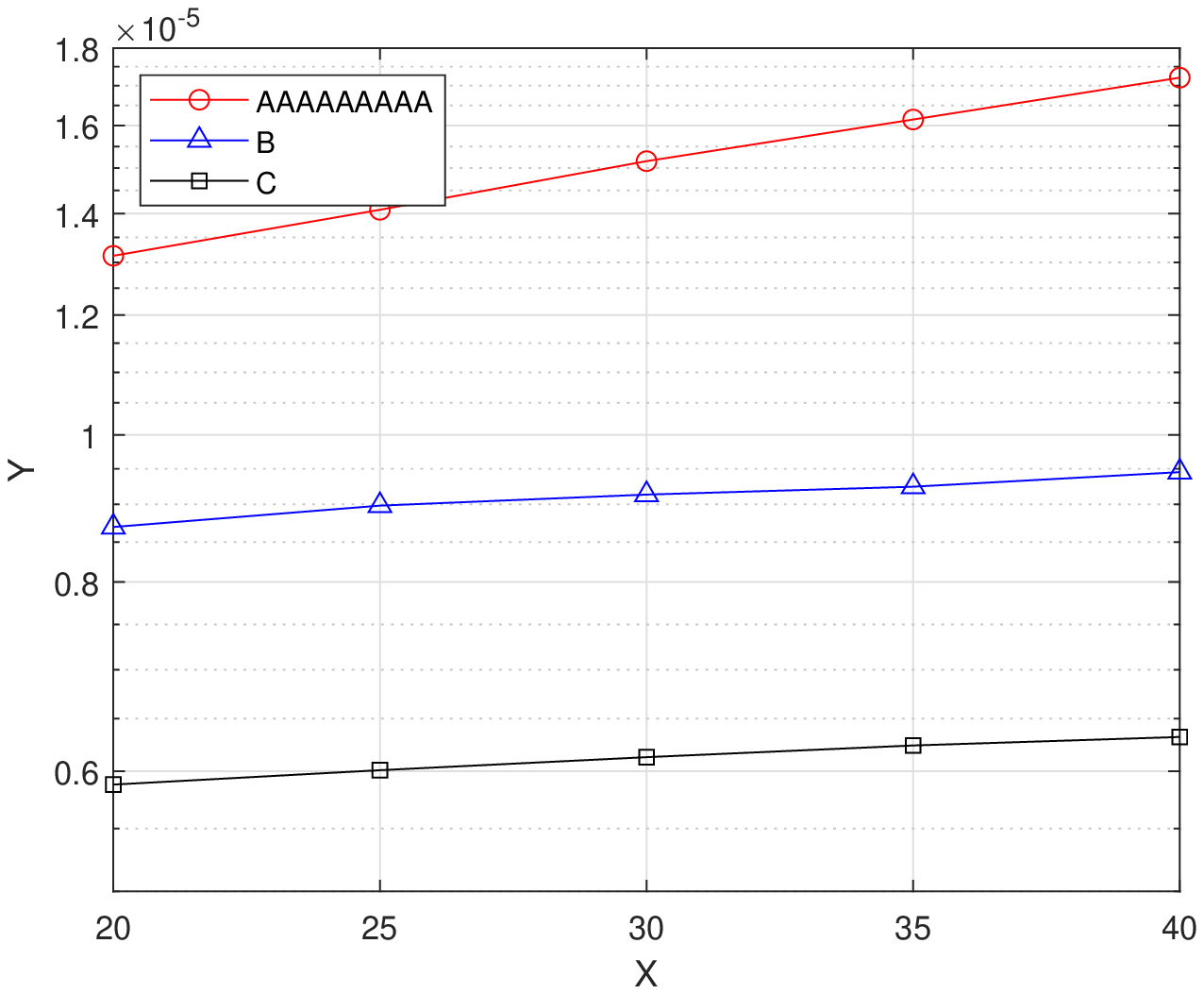}
		}
	\end{center}
	\vspace{-3.0mm}
	\caption{
		Critical \ac{BS} density $\lambdaBS^\text{c}$ according to $\hBS$ with different values $\hAU$
	}
	\label{fig:ESIS_crosspoint2}
\end{figure}

In Corollary \ref{co:col2} and Fig.~\ref{fig:ESIS_userratio_pout}, we show that when $\lambdaI$ is very small (i.e., noise-limited environments), the \ac{ES-BS} scheme outperforms the \ac{IS-BS} scheme for large $\lambdaBS$, but the \ac{IS-BS} scheme provides better performance than the \ac{ES-BS} scheme for small $\lambdaBS$.
Therefore, there exist the value of $\lambdaBS$ that makes the performance of two service provisioning schemes to be equal such as $\mathcal{P}^\text{IS}_\text{no}(\tiltIS)=\mathcal{P}^\text{ES}_\text{no}(\tiltgES,\tiltaES)$, and we define this value of $\lambdaBS$ as the \emph{critical density of \acp{BS}}, $\lambdaBS^\text{c}$.
That means in the region of $\lambdaBS<\lambdaBS^\text{c}$, the \ac{IS-BS} scheme is superior to the \ac{ES-BS} scheme in terms of the network outage probability and vice versa. 

Figure~\ref{fig:ESIS_crosspoint2} presents the critical density of the \ac{BS}, $\lambdaBS^\text{c}$,  as a function of the \ac{BS} height, $\hBS$, for the different values of the \ac{AU} height, $\hAU$.
In this figure, we can see that as the distance between the \ac{BS} and the \ac{AU} becomes closer, (i.e., $\hBS$ increases for given $\hAU$ or the $\hAU$ decreases for given $\hBS$), $\lambdaBS^\text{c}$ increases.
In this case, since the performance of the \acp{AU} is good enough due to the relatively short distance, the \ac{BS} in the \ac{IS-BS} scheme mainly tilt the antenna for \acp{GU} to enhance the network performance. Therefore, the \ac{IS-BS} scheme can provide better performance than the \ac{ES-BS} scheme.
In contrast, for the case that the \ac{BS} is far from the \acp{AU}, the \ac{BS} in the \ac{IS-BS} scheme has to properly tilt the antenna by considering the performance of both \ac{GU} and \ac{AU}. Therefore, in this case, the \ac{ES-BS} scheme can be more efficient as it can be independently optimized the antenna tilt angles for \acp{GU} and \acp{AU}, respectively.

\section{Conclusion}\label{sec:conclusion}
This paper explores an appropriate \ac{BS} service provisioning scheme to serve both \acp{GU} and \acp{AU} by considering tilt angle-based antenna gain.
We first derive the network outage probability for two types of provisioning schemes, i.e., \ac{IS-BS} scheme and \ac{ES-BS} scheme (in Theorem \ref{thm:netout}).
We then explore the conflict impact of the antenna tilt angle on the network outage probability, i.e., as the absolute value of the tilt angle decreases, the main lobe service area becomes wider, but the main link distance increases (in Corollary \ref{co:col1} and Remark \ref{re:rem1}).
From this relation, we numerically show that there exists the optimal \ac{BS} antenna tilt angle that minimizes the network outage probability.
Moreover, we show the impact of the ratio of \acp{GU}, the \ac{BS} height, the \ac{UAV} height, and densities of the total \acp{BS} and the interfering \acp{BS} on the optimal tilt angle as well as network outage probabilities for two service provisioning schemes.
Finally, for given network parameters, we present which service provisioning scheme is more appropriate.
Specifically, in Corollary \ref{co:col3}, we show that the \ac{ES-BS} scheme is better than the \ac{IS-BS} scheme when \acp{BS} are densely deployed. In contrast, the \ac{IS-BS} scheme performs better than the \ac{ES-BS} scheme for low \ac{BS} density or interfering \ac{BS} density. 
The outcomes of this work can be useful for the optimal antenna tilt angle design and the \ac{BS} provisioning service scheme determination in the networks, where both \acp{GU} and \acp{AU} exist.

\begin{appendix}
	\subsection{Proof of Theorem~\ref{thm:netout}} \label{app:thm1}
	For the given ratio of \acp{GU} and \acp{AU}, $\rho_\text{G}$ and $\rho_\text{A}$, the network outage probability can be presented by
	\begin{align} \label{eq:network_outage_SINR}
		\mathcal{P}_{\text{no}}^s\hspace{-0.4mm}(\hspace{-0.4mm}\tiltg\hspace{-0.3mm},\hspace{-0.4mm}\tilta\hspace{-0.4mm}) \hspace{-0.7mm}
		\mathop = \limits^{\mathrm{(a)}}\! \hspace{-0.7mm}
		\rho_{\text{G}} \mathcal{P}_{\text{no},\text{G}}^s\hspace{-0.4mm}(\hspace{-0.4mm}\tiltg\hspace{-0.4mm})\!
		+
		\!
		\rho_{\text{A}} \mathcal{P}_{\text{no},\text{A}}^s\hspace{-0.4mm}(\hspace{-0.4mm}\tilta\hspace{-0.4mm})
		,~\!s\!\in\!\hspace{-0.3mm}\{\text{IS},\hspace{-0.2mm}\text{ES}\},   \hspace{-1mm}
	\end{align}
	where $\mathcal{P}_{\text{no},i}^s(\tilti)$ is the network outage probability of $i$-type users and (a) is from the law of total probability.
	From \eqref{eq:SINR} and \eqref{eq:def_op_SINR},  
	$\pout{v}{\rvr}{\tilti}$ can be presented by
\begin{align}
	\pout{v}{\rvr}{\tilti}
	&= \label{eq:outage_thm_L}
	\mathbb{P}
	\left[ 
	\fading{\servingBS}
	<
	\frac{\target (\interference{s}{}+\noise)} {\tpower \ld{\rvr} 
	\hspace{-0.5mm}	
	 G_j(\rvr,\tilti) }
	\right] 
	\nonumber
	\\
	&
	\hspace{-6mm}
	\mathop = \limits^{\mathrm{(a)}} \hspace{0.5mm} 
	1
	\hspace{-0.5mm} - \hspace{-0.5mm}
	\mathbb{E}_{\interference{s}{}} \hspace{-1.5mm}
	\left[ 
	\frac
	{	\gamma\left( \mfactor{v}, \frac{\mfactor{v} \target (\interference{s}{}+\noise)} {\tpower \ld{\rvr}  G_j(\rvr,\tilti) }\right)	}
	{\Gamma(\mfactor{v})}
	\right]
	\nonumber
	\\
	&
	\hspace{-6mm}
	\mathop = \limits^{\mathrm{(b)}} \hspace{-0.5mm} 
	1 
	\hspace{-0.5mm} - \hspace{-0.5mm}
	\mathbb{E}_{\interference{s}{}} \hspace{-1.5mm}
	\left[ 
	\sum_{n=0}^{\mfactor{v}-1} \hspace{-1mm}
	\frac{1}{n!} \hspace{-1.5mm}
\left( 
	\frac{\mfactor{v} \target (\interference{s}{}+\noise) }{\tpower \ld{\hspace{-0.5mm} \rvr \hspace{-0.5mm}} 
		\hspace{-0.5mm} G_j(\rvr, \hspace{-0.5mm} \tilti)}
	\hspace{-0.5mm}
\right)^{\hspace{-1.5mm}n}  \right.\nonumber
	\\
	&
	\hspace{-6mm}
	\left.\quad\times \hspace{-0.5mm}
	\exp \hspace{-1mm}
	\left( \hspace{-1mm}
	-\frac{\mfactor{v} \target (\interference{s}{}+\noise)}{\tpower \ld{\rvr} \hspace{-1mm} G_j(\rvr,\tilti)}
	\right) 
	\right],
	\end{align}
	where (a) is from the \ac{CDF} of the Gamma distribution, and (b) follows from the definition of the incomplete gamma function for integer values of $\mfactor{v}$.
	From \eqref{eq:outage_thm_L}, we obtain \eqref{eq:gen_outage_net} by using 
	$\mathbb{E}_{I^s}[\exp(-z(I^s+\noise))]
	=
	\mathcal{L}_{I^s}(z) \exp(-z\noise)$ and 
	following property 
	\begin{align}
		\mathbb{E}_{I^s}[(-I^s)^n\exp(-zI^s)]
		=
		\frac{d}{dz^n}\mathcal{L}_{I^s}(z).
	\end{align}
	In \eqref{eq:gen_outage_net},  $\laplace{\interference{\text{IS}}{}}=\laplace{\interference{\text{IS}}{\text{0}}}$ and $\laplace{\interference{\text{ES}}{}}=\laplace{\interference{\text{ES}}{\text{G}}}\laplace{\interference{\text{ES}}{\text{A}}}$, 
	and $\laplace{\interference{s}{l}}$ is given by
	\begin{align} \label{eq:proof_laplace}
		&
		\laplace{\interference{s}{l}}	
		= \hspace{-0.5mm}
		\mathbb{E}_{\hpppI{l}} \hspace{-1.0mm}
		\left[ 
		\exp \hspace{-0.8mm}
		\left( \hspace{-1mm} -z \hspace{-4.mm}
		\sum_{\mathbf{x} \in \hpppI{l} \backslash  \{ \servingBS \}}^{} 
		\hspace{-5.0mm} \tpower \fading{\mathbf{x}} \ld{\hd} G_j(\hd,\tiltl)
		\hspace{-1.0mm}	\right) \hspace{-1.0mm}
		\right]
		\nonumber
		\\
		&\hspace{-0.5mm}= \hspace{-0.5mm}
		\mathbb{E}_{\hpppI{l}} \!\! \hspace{-0.5mm}
		\left[
		\prod_{\mathbf{x} \in \hpppI{l} \hspace{-0.3mm} \backslash \hspace{-0.3mm} \{\servingBS \hspace{-0.5mm}\}} \!\!\!\!\!\!
		\mathbb{E}_{\fading{\mathbf{x}} } \hspace{-0.8mm}
		\left[ 
		\exp \hspace{-0.5mm}
		\left\{ 
		-z \tpower \fading{\mathbf{x}}  \ld{\hd}
		 \hspace{-0.8mm}
		 G_j\hspace{-0.3mm}(\hspace{-0.3mm}\hd\hspace{-0.3mm},\hspace{-0.3mm}
		 \tiltl\hspace{-0.3mm}) \hspace{-0.1mm}
		\right\} \hspace{-0.5mm}
		\right] \hspace{-0.5mm}
		\right]
		\nonumber
		\\
		& \hspace{-0.5mm} \mathop = \limits^{\mathrm{(a)}} \hspace{-0.5mm}
		\mathbb{E}_{\hpppI{l}}  \!\! \hspace{-0.5mm}
		\left[
		\prod_{\mathbf{x} \in \hpppI{l} \backslash  \{\servingBS \}}
		\left\lbrace 
		\frac
		{	\losP{\text{L}}{\hd}	}
		{\left( 
			1+\frac{z}{\mfactor{\text{L}}} \tpower l_\text{L}(\hd) G_j(\hd,\tiltl)
			\right)^{\mfactor{\text{L}}}}	 
		\right. \right. 
		\nonumber
		\\
		&\left. \left.
		\hspace{13.0mm} +
		\frac{\losP{\text{N}}{\hd}}
		{1+z \tpower l_\text{N}(\hd) G_j(\hd,\tiltl)}	 
		\right\rbrace 
		\right],
	\end{align}
	where $\servingBS$ is the location of the serving \ac{BS}, and (a) is from the Laplace transforms of the Gamma distribution and the exponential distribution.
	From \eqref{eq:proof_laplace}, by applying the \ac{PGFL} \cite{HaeGan:09}, we obtain \eqref{eq:thm_laplace}.
	By averaging $\pout{v}{\rvr}{\tilti}$ over $\rvr$, $\mathcal{P}^s_{\text{no},i}(\tilti)$ in \eqref{eq:network_outage_SINR} is obtained as
	\begin{align}
		\mathcal{P}_{\text{no},i}^s(\tilti)
		&=
		\mathbb{E}_{\rvr}
		\left[ 
		\pout{v}{\rvr}{\tilti}
		\right] \!\!\!
		\nonumber
		\\
		&\mathop = \limits^{\mathrm{(a)}} \hspace{-3mm}
		\sum_{\substack{ v\in\{\text{L},\text{N}\},	\\	j\in\{1,2,3 \}  }	}^{} \hspace{-2mm}
		\left( 
		\int_{b_{k\hspace{-0.2mm} ,\hspace{-0.2mm} j \hspace{-0.3mm}}(\tilti)}
		^
		{b_{k\hspace{-0.2mm} ,\hspace{-0.2mm} j+\hspace{-0.2mm} 1 \hspace{-0.5mm}}(\tilti)}
		\mathcal{A}_{vj}^{a}
		\mathcal{P}^v_{\text{o},j}(r, \tilti) 
		f^{s,a}_{\rvr}(r)	dr	
		\right) 
		,
		\label{eq:outage_integral}
	\end{align}
	where (a) is from the definition of $\hpppBS{l}{vj}$ in \eqref{eq:layer}. 
	In \eqref{eq:outage_integral} $f^{s,a}_{\rvr}(r)=f_{r_i^{vj}}^{s,a}(r)$, and $b_{k, j}(\tilti)=b_{i,j}(\tilti)$ as $k\in\mathcal{U}_i$.
	By substituting \eqref{eq:outage_integral} into \eqref{eq:network_outage_SINR}, we obtain \eqref{eq:Network_Pout_general}.
\end{appendix}

\bibliographystyle{IEEEtran}
%

\bibliography{Bib/IEEEabrv,Bib/ISCGroup,Bib/StringDefinitions,Bib/bib_sj}

\begin{thebibliography}{10}
\providecommand{\url}[1]{#1}
\csname url@samestyle\endcsname
\providecommand{\newblock}{\relax}
\providecommand{\bibinfo}[2]{#2}
\providecommand{\BIBentrySTDinterwordspacing}{\spaceskip=0pt\relax}
\providecommand{\BIBentryALTinterwordstretchfactor}{4}
\providecommand{\BIBentryALTinterwordspacing}{\spaceskip=\fontdimen2\font plus
\BIBentryALTinterwordstretchfactor\fontdimen3\font minus
  \fontdimen4\font\relax}
\providecommand{\BIBforeignlanguage}[2]{{%
\expandafter\ifx\csname l@#1\endcsname\relax
\typeout{** WARNING: IEEEtran.bst: No hyphenation pattern has been}%
\typeout{** loaded for the language `#1'. Using the pattern for}%
\typeout{** the default language instead.}%
\else
\language=\csname l@#1\endcsname
\fi
#2}}
\providecommand{\BIBdecl}{\relax}
\BIBdecl

\bibitem{KimKimLee:20}
S.~Kim, M.~Kim, J.~Y. Ryu, and J.~lee, ``Impact of base station antenna tilt
  angle on {UAV} communications,'' in \emph{Proc. IEEE Global Commun. Conf.
  (GLOBECOM)}, Taipei, Taiwan, Dec. 2020, pp. {1--6}.

\bibitem{RinFedHel:20}
F.~Rinaldi, H.-L. Maattanen, J.~Torsner, S.~Pizzi, S.~Andreev, A.~Iera,
  Y.~Koucheryavy, and G.~Araniti, ``Non-terrestrial networks in {5G} \& beyond:
  A survey,'' \emph{IEEE Access}, vol.~8, pp. 165\,178--165\,200, Sep. 2020.

\bibitem{GioZor:21}
M.~Giordani and M.~Zorzi, ``Non-terrestrial networks in the {6G} era:
  Challenges and opportunities,'' \emph{{IEEE} Netw.}, vol.~35, no.~2, pp.
  244--251, Mar. 2021.

\bibitem{XinSteSeb:21}
\BIBentryALTinterwordspacing
X.~Lin, S.~Rommer, S.~Euler, E.~A. Yavuz, and R.~S. Karlsson, ``{5G} from
  space: An overview of {3GPP} non-terrestrial networks,'' \emph{{arXiv}}, Mar.
  2021. [Online]. Available: \url{https://arxiv.org/abs/2103.09156}
\BIBentrySTDinterwordspacing

\bibitem{YonRuiTen:16b}
Y.~Zeng, R.~Zhang, and T.~J. Lim, ``Wireless communications with unmanned
  aerial vehicles: opportunities and challenges,'' \emph{{IEEE} Commun. Mag.},
  vol.~54, no.~5, pp. 36--42, May 2016.

\bibitem{NasMotMil:17}
N.~H. Motlagh, M.~Bagaa, and T.~Taleb, ``{UAV}-based {IoT} platform: A crowd
  surveillance use case,'' \emph{{IEEE} Commun. Mag.}, vol.~55, no.~2, pp.
  128--134, Feb. 2017.

\bibitem{SamEvsRah:16}
S.~Hayat, E.~Yanmaz, and R.~Muzaffar, ``Survey on unmanned aerial vehicle
  networks for civil application: A communications viewpoint,'' \emph{{IEEE}
  Commun. Surveys Tuts.}, vol.~18, no.~4, pp. 2624--2661, Fourth Quart. 2016.

\bibitem{AziYunNan:18}
A.~A. Khuwaja, Y.~Chen, N.~Zhao, M.~Alouini, and P.~Dobbins, ``A survey of
  channel modeling for {UAV} communications,'' \emph{{IEEE} Commun. Surveys
  Tuts.}, vol.~20, no.~4, pp. 2804--2821, Fourth Quart. 2018.

\bibitem{AkrKar:18}
A.~{Al-Hourani} and K.~{Gomez}, ``Modeling cellular-to-{UAV} path-loss for
  suburban environments,'' \emph{IEEE Wireless Commun. Lett.}, vol.~7, no.~1,
  pp. {82--85}, Feb. 2018.

\bibitem{AlhKanLar:14}
A.~Al-Hourani, S.~Kandeepan, and S.~Lardner, ``Optimal {LAP} altitude for
  maximum coverage,'' \emph{IEEE Wireless Commun. Lett.}, vol.~3, no.~6, pp.
  {569--572}, Dec. 2014.

\bibitem{ChoLiuLee:18}
H.~Cho, C.~Liu, J.~Lee, T.~Noh, and T.~Q.~S. Quek, ``Impact of elevated base
  stations on the ultra-dense networks,'' \emph{{IEEE} Commun. Lett.}, vol.~22,
  no.~6, pp. 1268--1271, Jun. 2018.

\bibitem{AzaRosChe:18}
M.~M. Azari, F.~Rosas, K.~{Chen}, and S.~Pollin, ``Ultra reliable {UAV}
  communication using altitude and cooperation diversity,'' \emph{{IEEE} Trans.
  Commun.}, vol.~66, no.~1, pp. {330--344}, Jan. 2018.

\bibitem{PanDon:17}
P.~K. Sharma and D.~I. Kim, ``{UAV}-enabled downlink wireless system with
  non-orthogonal multiple access,'' in \emph{Proc. IEEE Global Commun. Conf.
  Workshops. (GC Wkshps)}, Singapore, Dec. 2017, pp. {1--6}.

\bibitem{HaiShuYon:18}
H.~{He}, S.~{Zhang}, Y.~{Zeng}, and R.~{Zhang}, ``Joint altitude and beamwidth
  optimization for {UAV}-enabled multiuser communications,'' \emph{{IEEE}
  Commun. Lett.}, vol.~22, no.~2, pp. {344--347}, Feb. 2018.

\bibitem{MohWalMer:16}
M.~{Mozaffari}, W.~{Saad}, M.~{Bennis}, and M.~{Debbah}, ``Efficient deployment
  of multiple unmanned aerial vehicles for optimal wireless coverage,''
  \emph{{IEEE} Commun. Lett.}, vol.~20, no.~8, pp. 1647--1650, Aug. 2016.

\bibitem{JiaYonRui:18}
J.~Lyu, Y.~Zeng, and R.~Zhang, ``{UAV}-aided offloading for cellular hotspot,''
  \emph{{IEEE} Trans. Wireless Commun.}, vol.~17, no.~6, pp. 3988--4001, Jun.
  2018.

\bibitem{ShoYonRui:19}
S.~Zhang, Y.~Zeng, and R.~Zhang, ``Cellular-enabled {UAV} communication: A
  connectivity-constrained trajectory optimization perspective,'' \emph{{IEEE}
  Trans. Commun.}, vol.~67, no.~3, pp. {2580--2604}, Mar. 2019.

\bibitem{KimLee:19}
M.~Kim and J.~Lee, ``Impact of an interfering node on unmanned aerial vehicle
  communications,'' \emph{{IEEE} Trans. Veh. Technol.}, vol.~68, no.~12, pp.
  12\,150--12\,163, Dec. 2019.

\bibitem{KimLeeQue:20}
D.~Kim, J.~Lee, and T.~Q.~S. Quek, ``Multi-layer unmanned aerial vehicle
  networks: Modeling and performance analysis,'' \emph{{IEEE} Trans. Wireless
  Commun.}, vol.~19, no.~1, pp. 325--339, Jan. 2020.

\bibitem{YonJiaRui:19}
Y.~Zeng, J.~Lyu, and R.~Zhang, ``Cellular-connected {UAV}: Potential,
  challenges, and promising technologies,'' \emph{{IEEE} Wireless Commun.},
  vol.~26, no.~1, pp. 120--127, Feb. 2019.

\bibitem{XinVijSiv:18}
X.~{Lin}, V.~{Yajnanarayana}, S.~D. {Muruganathan}, S.~{Gao}, H.~{Asplund},
  H.~{Maattanen}, M.~{Bergstrom}, S.~{Euler}, and Y.~.~E. {Wang}, ``The sky is
  not the limit: {LTE} for unmanned aerial vehicles,'' \emph{{IEEE} Commun.
  Mag.}, vol.~56, no.~4, pp. 204--210, Apr. 2018.

\bibitem{GalKibDas:18}
B.~Galkin, J.~Kibilda, and L.~A. DaSilva, ``Backhaul for low-altitude {UAVs} in
  urban environments,'' in \emph{Proc. IEEE Int. Conf. Commun. (ICC)}, Kansas
  City, MO, May 2018, pp. {1--6}.

\bibitem{AzaRosPol:17}
M.~M. Azari, F.~Rosas, A.~Chiumento, and S.~Pollin, ``Coexistence of
  terrestrial and aerial users in cellular networks,'' in \emph{Proc. IEEE
  Global Commun. Conf. Workshops. (GC Wkshps)}, Singapore, Dec. 2017, pp.
  {1--6}.

\bibitem{AzaRosPol:18}
M.~M. Azari, F.~Rosas, and S.~Pollin, ``Reshaping cellular networks for the
  sky: Major factors and feasibility,'' in \emph{Proc. IEEE Int. Conf. Commun.
  (ICC)}, Kansas City, MO, May 2018, pp. {1--7}.

\bibitem{RamWalNic:19}
R.~Amer, W.~Saad, and N.~Marchetti, ``Toward a connected sky: Performance of
  beamforming with down-tilted antennas for ground and {UAV} user
  co-existence,'' \emph{{IEEE} Commun. Lett.}, vol.~23, no.~10, pp. 1840--1844,
  Oct. 2019.

\bibitem{XiaYon:19}
X.~Xu and Y.~Zeng, ``Cellular-connected {UAV}: Performance analysis with {3D}
  antenna modeling,'' in \emph{Proc. IEEE Int. Conf. Commun. Workshops. (ICC
  Wkshps)}, Shanghai, China, May 2019, pp. {1--6}.

\bibitem{RamWalBor:20}
R.~Amer, W.~Saad, B.~Galkin, and N.~Marchetti, ``Performance analysis of mobile
  cellular-connected drones under practical antenna configurations,'' in
  \emph{Proc. IEEE Int. Conf. Commun. (ICC)}, Dublin, Ireland, Jun. 2020, pp.
  {1--7}.

\bibitem{RamWalNic:20}
R.~Amer, W.~Saad, and N.~Marchetti, ``Mobility in the sky: Performance and
  mobility analysis for cellular-connected {UAVs},'' \emph{{IEEE} Trans.
  Commun.}, vol.~68, no.~5, pp. 3229--3246, May 2020.

\bibitem{3GPP:TR:36.814:V9.2.0}
3rd {G}eneration~{P}artnership {P}roject, ``Technical specification group radio
  access network; evolved universal terrestrial radio access ({E-UTRA});
  further advancements for {E-UTRA} physical layer aspects,'' TR 36.814 V9.2.0,
  Tech. Rep., Mar. 2017, release 9.

\bibitem{ZhuZheFit:18}
Y.~Zhu, G.~Zheng, and M.~Fitch, ``Secrecy rate analysis of uav-enabled mmwave
  networks using matérn hardcore point processes,'' \emph{{IEEE} J. Sel. Areas
  Commun.}, vol.~36, no.~7, pp. 1397--1409, Jul. 2018.

\bibitem{LyuWan:21}
J.~Lyu and H.-M. Wang, ``Secure uav random networks with minimum safety
  distance,'' \emph{{IEEE} Trans. Veh. Technol.}, vol.~70, no.~3, pp.
  2856--2861, Mar. 2021.

\bibitem{ZhiLaiGua:18}
Z.~Yang, L.~Zhou, G.~Zhao, and S.~Zhou, ``Blockage modeling for inter-layer
  {UAVs} communications in urban environments,'' in \emph{Proc. IEEE Int. Conf.
  Telecommun. (ICT)}, St. Malo, France, Jun. 2018, pp. {307--311}.

\bibitem{HerZaiMcl:15}
R.~Hernandez-Aquino, S.~A.~R. Zaidi, D.~McLernon, M.~Ghogho, and A.~Imran,
  ``Tilt angle optimization in two-tier cellular networks—a stochastic
  geometry approach,'' \emph{{IEEE} Trans. Commun.}, vol.~63, no.~12, pp.
  {5162--5177}, Dec. 2015.

\bibitem{AhnLee:14}
S.~M. {Razavizadeh}, M.~{Ahn}, and I.~{Lee}, ``Three-dimensional beamforming: A
  new enabling technology for 5g wireless networks,'' \emph{{IEEE} Signal
  Process. Mag.}, vol.~31, no.~6, pp. 94--101, Nov. 2014.

\bibitem{MohHal:18}
M.~Alzenad and H.~Yanikomeroglu, ``Coverage and rate analysis for unmanned
  aerial vehicle base stations with {LoS/NLoS} propagation,'' in \emph{Proc.
  IEEE Global Commun. Conf. Workshops. (GC Wkshps)}, Abu Dhabi, UAE, Dec. 2018,
  pp. {1--7}.

\bibitem{BacBla:09b}
F.~Baccelli and B.~B{\l}aszczyszyn, \emph{Stochastic Geometry and Wireless
  Networks, Volume II --- Applications}, ser. Foundations and Trends in
  Networking.\hskip 1em plus 0.5em minus 0.4em\relax NoW Publishers, 2009.

\bibitem{HesAhmMoh:16}
H.~{ElSawy}, A.~{Sultan-Salem}, M.~{Alouini}, and M.~Z. {Win}, ``Modeling and
  analysis of cellular networks using stochastic geometry: A tutorial,''
  \emph{{IEEE} Commun. Surveys Tuts.}, vol.~19, no.~1, pp. 167--203, First
  Quart. 2017.

\bibitem{SalIkk:09}
S.~S. {Ikki} and M.~H. {Ahmed}, ``Performance analysis of decode-and-forward
  incremental relaying cooperative-diversity networks over rayleigh fading
  channels,'' in \emph{Proc. IEEE Veh. Technol. Conf. (VTC)}, Barcelona, Spain,
  Apr. 2009, pp. 1--6.

\bibitem{tableint:07}
I.~S. Gradshteyn and I.~M. Ryzhik, \emph{Table of Integrals, Series, and
  Products}, 7th~ed.\hskip 1em plus 0.5em minus 0.4em\relax San Diego, CA, USA:
  Academic Press, 2007.

\bibitem{3GPP:TR:36.777:V15.0.0}
3rd {G}eneration~{P}artnership {P}roject, ``Technical specification group radio
  access network; study on enhanced lte support for aerial vehicles,'' TR
  36.777 V15.0.0, Tech. Rep., Dec. 2017, release 15.

\bibitem{HolPec:08}
J.~Holis and P.~Pechac, ``Elevation dependent shadowing model for mobile
  communications via high altitude platforms in built-up areas,'' \emph{{IEEE}
  Trans. Antennas Propag.}, vol.~56, no.~4, pp. 1078--1084, Apr. 2008.

\bibitem{HaeGan:09}
M.~Haenggi and R.~K. Ganti, ``Interference in large wireless networks,''
  \emph{Foundations and Trends in Networking}, vol.~3, no.~2, pp. 127--248,
  2009.

\end{thebibliography}

\end{document}